%% file: 00_main.tex
\documentclass{lmcs}
\pdfoutput=1

\usepackage{lastpage}
\lmcsdoi{22}{4}{17}
\lmcsheading{}{\pageref{LastPage}}{}{}%
{Nov.~05,~2024}{Nov.~05,~2026}{}

\input{macros}

\begin{document}

\title{Explorable Parity Automata}
\author[E. Hazard]{Emile Hazard}[a]
\author[O. Idir]{Olivier Idir\lmcsorcid{https://orcid.org/0009-0003-3848-8515}}[b]
\author[D. Kuperberg]{Denis Kuperberg\lmcsorcid{https://orcid.org/0000-0001-5406-717X}}[a]

\address{CNRS, LIP, ENS Lyon, France}
\email{emile.hazard@ens-lyon.fr}
\email{denis.kuperberg@ens-lyon.fr}
\address{IRIF, Université Paris-Cité, France}
\email{olivier.idir@irif.fr}

\input{01_intro}

\input{02_defs_and_props}

\input{02.5_expressivity}

\input{03_complexity}

\input{04_infinite_tokens}

\input{05_conclusion}

\bibliography{biblio}
\bibliographystyle{alphaurl}
\end{document}

%% file: 01_intro.tex


\begin{abstract}
    We define the class of explorable automata on finite or infinite words. This is a generalisation of History-Deterministic (HD) automata, where this time non-deterministic choices can be resolved by building finitely many simultaneous runs instead of just one.
    We show that recognizing HD parity automata of fixed index among explorable ones is in \PTIME, thereby giving a strong link between the two notions. We then show that recognizing explorable automata is \EXPTIME-complete, in the case of finite words or parity automata up to index $[0,2]$. Additionally, we define the notion of $\omega$-explorable automata on infinite words, where countably many runs can be used to resolve the non-deterministic choices. We show \EXPTIME-completeness for $\omega$-explorability of automata on infinite words for the safety and coB\"uchi acceptance conditions.
    We finally characterize the expressivity of ($\omega$-)explorable automata with respect to the parity index hierarchy.
\end{abstract}

\maketitle

\section{Introduction}

In several fields of theoretical science, the tension between deterministic and non-deterministic models is a source of fundamental open questions, and has led to important lines of research. The most famous of this kind is the P vs NP question in complexity theory. This paper aims at further investigating the frontier between determinism and non-determinism in automata theory. Although Non-deterministic and Deterministic Finite Automata (NFA and DFA) are known to be equivalent in terms of expressive power, many subtle questions remain about the cost of determinism, and a deep understanding of non-determinism will be needed to solve them.

One of the approaches investigating non-determinism in automata is the study of History-Deterministic (HD) automata, introduced in~\cite{GFG06} under the name Good-For-Games (GFG) automata. An automaton is HD if, when reading input letters one by one, its non-determinism can be resolved on-the-fly without any need to guess the future. This constitutes a model that is intermediary between non-determinism and determinism, and can sometimes bring the best of both worlds. Like deterministic automata, HD automata on infinite words retain good properties such as their soundness with respect to composition with games, making them appropriate for use in Church synthesis algorithms~\cite{GFG06}. On the other hand, like non-deterministic automata, they can be exponentially more succinct than deterministic ones~\cite{KS15}.
There is a very active line of research trying to understand the various properties of HD automata, see e.g.~\cite{AK22,BKLS20,BL22,PhDCasares23} for some of the recent developments.
The terminology \emph{history-deterministic}, was introduced originally in the theory of regular cost functions~\cite{Col09}. The name ``history-deterministic'' corresponds to the above intuition of solving non-determinism on-the-fly, while the earlier name of  ``good-for-games'' refers to sound composition with games. These two notions may actually differ in some quantitative frameworks, but coincide on Boolean automata~\cite{BL21}, and have been used interchangeably in most of the literature on the topic. In this paper, since we are mainly interested in resolving the non-determinism on-the-fly, we choose the HD denomination to emphasize this aspect\footnote{This departs from earlier practices consisting in using HD and GFG in a way coherent with their contexts of introduction: HD for cost functions and GFG for Boolean automata. Hence most of the papers cited here use GFG.}.

The goal of this paper is to pursue this line of research by introducing and studying the class of explorable automata on finite and infinite words.
The intuition behind explorability is to limit the amount of non-determinism required by the automaton to accept its language, in a more permissive way than HD automata. If $k\in\N$, an automaton is $k$-explorable if when reading input letters, it suffices to keep track of $k$ runs to build an accepting one, if it exists. An automaton is explorable if it is $k$-explorable for some $k\in\N$.
This can be seen as a variation on the notion of HD automaton, which corresponds to the case $k=1$.
The present work can be compared to~\cite{KM19}, where a notion related to $k$-explorability (called \emph{width}) is introduced and studied, see Section \ref{subsec:explodef}. In particular, some results of~\cite{KM19} also apply to $k$-explorability, notably \EXPTIME-completeness of deciding $k$-explorability of an NFA if $k$ is part of the input. Surprisingly however, the techniques used in~\cite{KM19} are quite different from the ones we need here. This shows that fixing a bound $k$ for the number of runs leads to very different problems compared to asking for the existence of such a bound.

We then proceed to study the decidability and complexity of the explorability problem: deciding whether an input automaton on finite or infinite words is explorable.
For this, we establish a connection with the population control problem studied in~\cite{Bertrand}. This problem asks, given an NFA with an arbitrary number of tokens in the initial state, whether a controller can choose input letters, thereby forcing every token to reach a designated state, even if tokens are controlled by an opponent.
It is shown in~\cite{Bertrand} that the population control problem is \EXPTIME-complete, and we adapt their proof to our setting to show that the explorability problem is \EXPTIME-complete as well, already for NFAs.  We also show that a direct reduction is possible, but at an exponential cost, yielding only a $2$-\EXPTIME algorithm for the NFA explorability problem. In the case of infinite words, we adapt the proof to $[0,2]$-parity case, thereby showing that the $[0,2]$-explorability problem is in \EXPTIME as well.
We also remark that, as in~\cite{Bertrand}, the number of tokens needed to witness explorability can go as high as doubly exponential in the size of the automaton.


Notice that interestingly, from a model-checking perspective, our approach is dual to~\cite{Bertrand}: in the population control problem, an NFA is well-behaved when we can ``control'' it by forcing arbitrarily many runs to simultaneously reach a designated state, via an appropriate choice of input letters. On the contrary, in our approach, the input letters form an adversarial environment, and our NFA is well-behaved when its non-determinism is limited, in the sense that it is enough to spread finitely many runs to explore all possible behaviors.

We also establish the expressivity of explorable automata, through reduction to or from deterministic parity automata. Surprisingly, the expressivity hierarchy of the explorable automata collapse, just as the hierarchy of non-deterministic automata, albeit one step later. The general case is reached with $[1,3]$-parity explorable automata, which can recognize all regular languages.

On infinite words, we push further the notion of explorability, by remarking that for some automata, even following a countably infinite number of runs is not enough. This leads to defining the class of \emph{$\omega$-explorable automata}, as those automata on infinite words where non-determinism can be resolved using countably many runs.
We show that $\omega$-explorable automata form a non-trivial class even for the safety acceptance condition (but not for reachability), and give an \EXPTIME algorithm recognizing $\omega$-explorable automata, encompassing the safety and coB\"uchi conditions. We also show \EXPTIME-hardness of this problem, by adapting the \EXPTIME-hardness proof of~\cite{Bertrand} to the setting of $\omega$-explorability.
\smallskip

\textbf{Summary of the contributions.}

We show that given an NFA or a parity automaton with parities $\subseteq [0,2]$, it is decidable and \EXPTIME-complete to check whether it is explorable. We also study the expressivity in terms of recognized languages of the different parity classes of explorable automata.
Our proof of \EXPTIME-completeness for NFA explorability uses techniques developed in~\cite{Bertrand}, where \EXPTIME-completeness is shown for the NFA population control problem. We generalise this result to \EXPTIME explorability checking for $[0,2]$-parity automata, requiring further adaptations. We also give a black box reduction using the result from~\cite{Bertrand}. This is enough to show decidability of the NFA explorability problem, but it yields a $2$-\EXPTIME algorithm. As in~\cite{Bertrand}, the \EXPTIME algorithm yields a doubly exponential tight upper bound on the number of tokens needed to witness explorability. We show that deciding the explorability of $[1,3]$-parity automata amounts to deciding the explorability of automata in the general parity case.

On infinite words, we show that any reachability automaton is $\omega$-explorable, but that this is not the case for safety automata.
We show that both the safety and coB\"uchi $\omega$-explorability problems are \EXPTIME-complete.
We also show that the \Buchi case corresponds to the general case : any nondeterministic parity automaton can be converted in \PTIME to a \Buchi automaton with same $\omega$-explorability status.
\smallskip

\textbf{Related Works.}

\noindent Zooming out, many works aim at quantifying the amount of non-determinism in automata.
A survey by Colcombet~\cite{Col12} gives useful references on this question.
Let us mention for instance the notion of ambiguity, which quantifies the number of simultaneous accepting runs. Similarly to~\cite{KM19}, we can note that ambiguity is orthogonal to $k$-explorability. Remark however that our finite/countable/uncountable explorability hierarchy is reminiscent of the finite/polynomial/exponential ambiguity hierarchy (see \eg~\cite{Amb}).

In~\cite{nondet}, several ways of quantifying the non-determinism in automata are studied from the point of view of complexity, including notions such as the number of advice bits needed.

Another approach is studied in~\cite{PSA17}, where a measure of the maximum non-deterministic branching along a run is defined and compared to other existing measures.

Following the HD approach, a hierarchy of non-determinism and an analysis of this hierarchy via probabilistic models is given in~\cite{AKL21}.

\noindent The particular problem of deciding in polynomial time whether a given automaton is HD has attracted a lot of attention. After some initial results on NFA, B\"uchi and coB\"uchi conditions~\cite{LR13,KS15,BK18,BKLS20}, the general result was recently obtained by Lehtinen and Prakash in~\cite{LP25}, showing that for any fixed parity index (and consequently for any fixed $\omega$-regular condition) it is decidable in polynomial time whether a given parity automaton is HD. It is shown in \cite{LP25} that the two-token game introduced in~\cite{BK18} for B\"uchi automata actually characterizes HDness for any acceptance condition, as conjectured in~\cite{BK18}.
The inspiration for $k$-explorability stems from this kind of approach, where the idea of following a finite number of runs in parallel plays a central role. Remark however that the notion of explorability as studied here is stronger than what is needed in the two-token game from \cite{BK18,LP25}. 
The $k$-explorability (and explorability) property was explicitly defined under the name $k$-History-Determinism in~\cite{BL22}, as a proof tool to decide the HDness of LimInf and LimSup automata. The work~\cite{BL22} is part of a research effort to understand how partial determinism notions such as HDness play out in quantitative automata, see survey~\cite{Boker22}.
Our goal here is to investigate explorability as defining a natural class of automata on finite and infinite words, somehow giving it an ``official status'' not restricted to an intermediate proof tool.


\smallskip

\textbf{History of this work.}
\noindent It is traditional in our community to present results as a finished product, abstracting away the path that led to it. This paragraph is an experiment: we believe that in addition to this practice, it can be interesting for the reader to have access to a history of how ideas developed.

The interest we took in the explorability notion originated in the fact that it makes deciding HDness much easier, and the hope was that by using this notion as an intermediate, we could obtain an algorithm improving on the \EXPTIME upper bound for deciding HDness of parity automata of fixed index, e.g.~to \PSPACE. As we described above, we ended up showing that this approach cannot yield an algorithm below \EXPTIME (at least not in full generality). However, although this was initially only a tool for this decision problem, explorability turns out to be a natural generalisation of HD automata, and an interesting class to study in itself. The first investigation of this notion, and in particular of its decidability, was the object of a short research internship by Milla Valnet under the supervision of the third author. It was expected that decidability of explorability would be a reachable goal for such a short internship, but it turned out that this was overly optimistic. The internship yielded preliminary results, and in particular was useful to introduce and study the notion of ``coverability''.  This version of the problem does not take acceptance conditions into account, but only asks that at any point of the run, every state that could be reached is actually occupied by a token. After the internship, we continued to use this coverability notion as a stepping stone towards an understanding of explorability. However, after more preliminary results and unsuccessful attempts at obtaining decidability, we discovered the connection between explorability and population control from~\cite{Bertrand}, that rendered the intermediate notion of coverability useless for our purposes, and we then focused on exploiting that link. We chose to leave coverability out of the present exposition, as it feels like a ``watered-down'' version of explorability, but it could be useful in some contexts, hence we briefly mention it in this chronological account. Let us just informally state here that it is straightforward to modify our proofs in order to show that deciding whether an NFA is coverable is \EXPTIME-complete as well.
The first results were obtained during the PhD of Emile Hazard, and published in~\cite{HK23}. Some new results, namely \EXPTIME algorithm for coB\"uchi and $[0,2]$-explorability, and expressivity results, were obtained during the internship of Olivier Idir. This paper, extending~\cite{HK23}, aims at gathering what we currently know about explorable automata.

%% file: 02_defs_and_props.tex
\section{Explorable automata}

\subsection{Preliminaries}\label{sec:defs}
If $i\leq j$ are integers, we will denote by $[i,j]$ the integer interval $\{i,i+1,\dots,j\}$. If $S$ is a set, its cardinal will be denoted $|S|$, and its powerset $\P(S)$.
\subsection{Automata}

We work with a fixed finite alphabet $\Sigma$.
We will use the following default notation for the components of an automaton $\A$: $Q_\A$ for its set of states, $q_0^\A$ for its initial state, $F_\A$ for its accepting states, $\Delta_\A$ for its set of transitions. If the automaton is clear from context, the subscript/superscript $\A$ might be omitted. We might also specify its alphabet by $\Sigma_\A$ instead of $\Sigma$ for cases where different alphabets come into play.
If $\Delta\subseteq Q\times\Sigma\times Q$ is the transition relation, and $(p,a)\in Q\times\Sigma$, we will note  $\Delta(p,a)=\{q\in Q, (p,a,q)\in\Delta\}$. If $X\subseteq Q$, we note $\Delta(X,a)=\bigcup_{p\in X} \Delta(p,a)$. A transition $(p,a,q)$ will often be noted $p\trans{a}q$.

To simplify definitions, all automata in this paper will be assumed to be complete (by adding a rejecting sink state if needed). This means that for all $(p,a)\in Q\times\Sigma$, we assume $\Delta(p,a)\neq\emptyset$. The rejecting sink state will often be implicit in our constructions and examples.

We will consider non-deterministic automata on finite words (NFAs). A run of such an automaton on a word $a_1a_2\dots a_n\in\Sigma^*$ is a sequence of transitions $\delta_1\dots \delta_n\in \Delta^*$, such that there exists a sequence of states $q_0,\dots, q_n$ and for all $i\in[1,n], \delta_i=(q_{i-1},a_i,q_i)$, ($q_0$ being the initial state). Such a run is accepting if $q_n\in F$. As usual, the language of an automaton $\A$, denoted $L(\A)$, is the set of words that admit an accepting run.

We will also deal with automata on infinite words, and we recall here some of the standard acceptance conditions for such automata. A run on an infinite word $w=a_1a_2\dots\in\Sigma^\omega$ is now an infinite sequence of transitions $\delta_1,\delta_2,\dots$, \ie an element of $\Delta^\omega$. As before there must exist an underlying sequence of states $q_0,q_1,q_2,\dots$ with $q_0$ the initial state, such that for each $i\geq 1$, we have $\delta_i=(q_{i-1},a_i,q_i)$. 

The acceptance conditions safety, reachability, B\"uchi and coB\"uchi are defined with respect to an accepting subset of transitions $F\subseteq\Delta$. Here are the languages of accepting runs among $\Delta^\omega$, for these four acceptance conditions:
\begin{itemize}
    \item Safety: $F^\omega$
    \item Reachability: $\Delta^*F\Delta^\omega$
    \item B\"uchi: $(\Delta^*F)^\omega$
    \item coB\"uchi: $\Delta^*F^\omega$
\end{itemize}
Transitions from $F$ will be called B\"uchi transitions in B\"uchi automata, and transitions from $\Delta\setminus F$ will be called coB\"uchi transitions in coB\"uchi automata.

Finally, we will also use the parity acceptance condition: it uses a ranking function $\rk$ from $\Delta$ to an interval of integers $[i,j]$, called the \emph{parity index} of the automaton. A run is accepting if the maximal rank appearing infinitely often is even.

For conciseness, we will simply write \emph{$[i,j]$-automaton} for a parity automaton using ranks from $[i,j]$.
Remark that B\"uchi automata correspond to $[1,2]$-automata, and coB\"uchi automata to $[0,1]$-automata.

For all these acceptance conditions on infinite words, we will sometimes use state-based acceptance instead of transition-based when more convenient for our constructions. Recall that we can switch from transition-based to state-based via a state blow-up by a factor of number of priorities (e.g. doubling the number of states for B\"uchi or coB\"uchi condition), and the translation from state-based to transition-based can be done without changing the size of the automaton. These translations do not affect any of the explorability properties considered in this paper. See~\cite{PhDCasares23} for details on the merits of transition-based acceptance conditions over the state-based ones.

\subsection{Games}
%

A \emph{game} $\G=(V_0,V_1,v_I,E, W)$ of infinite duration between two players $0$ and $1$ consists of: a finite set of \emph{positions} $V$ being a disjoint union of $V_0$ and $V_1$; an \emph{initial position} $v_I\in V$; a set of \emph{edges} $E\subseteq V\times V$; and a \emph{winning condition} $W\subseteq V^\omega$. 
We will later use names more explicit than $0$ and $1$ for the players, describing their roles in the various games we will define.

A \emph{play} is an infinite sequence of positions $v_0v_1v_2\dots\in V^\omega$ such that $v_0=v_I$ and for all $n\in \mathbb N$, $(v_n,v_{n+1})\in E$. A play $\pi\in V^\omega$ is \emph{winning} for Player 0 if it belongs to $W$. Otherwise, $\pi$ is \emph{winning} for Player 1.

A \emph{strategy} for Player $0$ (resp. $1$) is a function ${\sigma_0}:{V^\ast\times V_0}\to{V}$ (resp. ${\sigma_1}:{V^\ast\times V_1}\to{V}$), describing which edge should be played given the history of the play $u\in V^\ast$ and the current position $v\in V$. A strategy $\sigma_P$ has to obey the edge relation, i.e.~there has to be an edge in $E$ from $v$ to $\sigma_P(u,v)$. A play $\pi=v_0v_1v_2\dots$ is \emph{consistent} with a strategy $\sigma_P$ of a player $P$ if for every $n$ such that $v_n\in V_P$ we have $v_{n+1}=\sigma_P(v_0\ldots v_{n-1},v_n)$.

A strategy for Player $0$ (resp. Player $1$) is \emph{positional} (or \emph{memoryless}) if it does not use the history of the play, i.e. it can be seen as a function $V_0\to V$ (resp. $V_1\to V$).

We say that a strategy $\sigma_P$ of a player $P$ is \emph{winning} if every play consistent with $\sigma_P$ is winning for $P$. In this case, we say that $P$ \emph{wins} the game $\G$.

A game is \emph{positionally determined} if one of the players has a positional winning strategy in the game.

See \eg~\cite{GTW02} for more details on games and strategies.

We could equivalently define a winning condition on edges: $W\subseteq E^\omega$, but we will not need to dive into this distinction here. This is because in the interest of readability, when describing games in the paper, we will not give explicit definitions of the sets $V_0$, $V_1$ and $E$, but give slightly more informal descriptions in terms of possible actions of players at each round. It is straightforward to build a formal description of the games from such a description.

\subsection{Explorability}\label{subsec:explodef}

We start by introducing the \emph{$k$-explorability game}, which is the central tool allowing us to define the class of explorable automata.

\begin{defi}[$k$-explorability game]\label{def:explo}
    Consider a non-deterministic automaton $\A$ on finite or infinite words, and an integer $k$. The $k$-explorability game on $\A$ is played on the arena $Q^k$. We will consider that a position $S=(q_1,\dots,q_k)$ of the arena $Q^k$ is described via $k$ tokens, numbered $1$ to $k$, and placed respectively in the states $q_1,\dots,q_k$ of $\A$. If $l\in[1,k]$ is the index of a token, we will note $S(l)=q_l$ the location of this token in position $S$. The two players are called \agents and \controller, and they play as follows.
    \begin{itemize}
        \item The initial position is the $k$-tuple $S_0=(q_0, \ldots, q_0)$.
        \item At step $i\geq 1$, from a position $S_{i-1}\in Q^k$, \controller chooses a letter $a_i\in\Sigma$, and \agents chooses $S_i\in Q^k$ such that for every token $l\in[1,k]$, $S_{i-1}(l) \labelarrow{a_i} S_i(l)$ is a transition of $\A$.
    \end{itemize}
    The play is won by \agents if for all $\beta \leq \omega$ such that the word $(a_i)_{1 \leq i < \beta}$ is in $\L(\A)$, there is a token $l\in[1,k]$ being accepted by $\A$, meaning that the sequence $(S_i(l))_{i < \beta}$ is an accepting run\footnote{This condition $\beta \leq \omega$ is actually accounting separately for the two cases of finite and infinite words, corresponding respectively to $\beta < \omega$ and $\beta = \omega$.}. Otherwise, the winner is \controller.\\
    We will say that $\A$ is \emph{$k$-explorable} if \agents wins the $k$-explorability game (\ie has a winning strategy, ensuring the win independently of the choices of \controller).\\
    We will say that $\A$ is \emph{explorable} if it is $k$-explorable for some $k\in\N$.
\end{defi}

\begin{figure}
    \centering
    \input{automata/Ak_nonexpl}
    \caption{An explorable and a non-explorable automata}
    \label{fig:ex_explo_non_explo}
\end{figure}
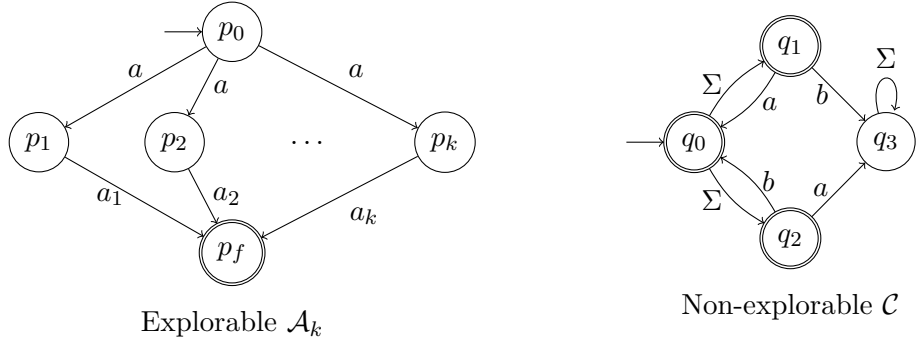

\begin{exa}\label{ex:explo}
Consider the automata from \Cref{fig:ex_explo_non_explo}. The NFA $\A_k$ on alphabet $\{a,a_1,\dots,a_k\}$ is $k$-explorable, but not $(k-1)$-explorable. It can easily be strengthened to a binary alphabet whilst still requiring $k$ tokens, by replacing the transition labels $a_1,\dots,a_k$ by distinct words of the same length.

On the other hand, the NFA $\C$ is a non-explorable NFA accepting all words on alphabet $\Sigma=\{a,b\}$. Indeed, \controller can win the $k$-explorability game for all $k$, by sending tokens to the sink $q_3$ one by one, choosing at each step the letter $b$ if $q_1$ is occupied by at least one token, and the letter $a$ otherwise.
\end{exa}

\begin{exa}\label{ex:exptokens}
The NFA $\B_k$ from \Cref{fig:exp_tokens} with $3k+1$ states on alphabet $\Sigma=\{a,b\}$ is explorable, but requires $2^k$ tokens. Indeed, since when choosing the $2i^\mathrm{th}$ letter \controller can always pick the state $p_i$ or $r_i$ containing the least amount of tokens to decide whether to play $a$ or $b$, the best strategy for \agents is to split his tokens evenly at each $q_i$. This means he needs to start with $2^k$ tokens to end up with at least one token in $q_k$ after a word of $\Sigma^{2k}$.

\begin{figure}
    \centering
    \input{automata/ex_split_exp}
    \caption{An explorable automaton $\B_k$ requiring exponentially many tokens}
    \label{fig:exp_tokens}
\end{figure}
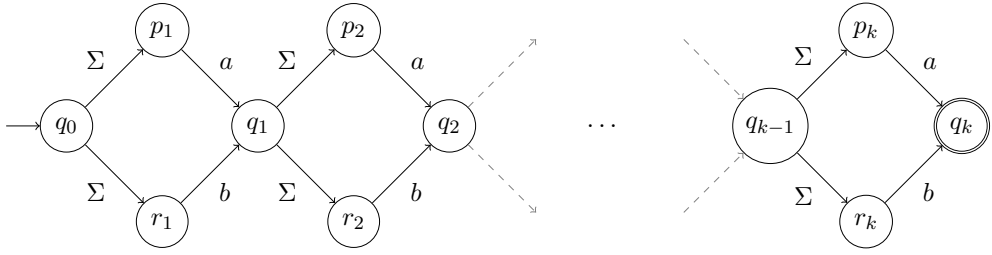
\end{exa}

Let us mention a few facts that follow from the definition of explorability:

\pagebreak

\begin{lem}
    ~
    \begin{itemize}
        \item Any automaton for a finite language (i.e. containing finitely many words) is explorable.
        \item If $\A$ is $k$-explorable, then it is $n$-explorable for all $n\geq k$.
        \item If $\A$ is $k$-explorable and $\B$ is $n$-explorable, then 
        \begin{itemize} 
            \item $\A\cup\B$ (with states $Q=\{q_0\}\cup Q_\A\cup Q_\B$) is $(k+n)$-explorable,
            \item the union product $\A\times\B$ (with $F=(F_\A\times Q_\B)\cup (Q_\A\times F_\B)$) is $\max(k,n)$-explorable, 
            \item the intersection product $\A\times\B$ (with $F=F_\A\times F_\B$) is $(kn)$-explorable.
        \end{itemize}
    \end{itemize}
\end{lem}

\begin{proof}
If $L(\A)$ is finite, it is enough to take $k=|L(\A)|$ tokens to witness explorability: for each $u\in L(\A)$, the token $t_u$ assumes that the input word is $u$ and follows an accepting run of $\A$ over $u$ as long as input letters are compatible with $u$. As soon as an input letter is not compatible with $u$, the token $t_u$ is discarded and behaves arbitrarily for the rest of the play.

If $\A$ is $k$-explorable and $n\geq k$, then \agents can win the $n$-explorability game by using the same strategy with the first $k$ tokens and making arbitrary choices with the $n-k$ remaining tokens.

If $\A$ and $\B$ are $k$- and $n$-explorable respectively, then \agents can use both strategies simultaneously with $k+n$ tokens in $\A\cup\B$, using $k$ tokens in $\A$ and $n$ tokens in $\B$. If the input word is in $\A$ (resp. $\B$), then the tokens playing in $\A$ (resp. $\B$) will win the play. 

In the union product $\A\times\B$, it is enough to take $\max(k,n)$ tokens: if $1\leq i\leq\min(k,n)$, the token number $i$ follows the strategy of the token $i$ in $\A$ on the first coordinate, and the strategy of the token $i$ in $\B$ in the second one. If $\min(k,n)< i\leq\max(k,n)$, say wlog $k< i\leq n$, the token $i$ follows an arbitrary strategy on the $\A$-component and the strategy of token $i$ on the $\B$-component.

However, \agents may need up to $kn$ tokens to play in $\A\times\B$ when the accepting set is $F_\A\times F_\B$, as \emph{one} token needs to accept both paths of the product. The token $(i,j)$ will use the strategy of the token $i$ in the $k$-explorability game of $\A$ together with the strategy of the token $j$ in the $n$-explorability game of $\B$. This lower bound of $kn$ cannot be improved: consider for instance the intersection product $\A_k\times\A_n$, where $\A_k,\A_n$ are from Example \ref{ex:explo}, using as alphabet the cartesian product of their respective alphabets: $\{a,a_1,a_2,\dots,a_k\}\times\{a,a_1,\dots,a_n\}$, or $\{a,b\}^2$ in their binary alphabet versions.
\end{proof}

\noindent
Notice that a similar notion was introduced in~\cite{KM19} under the name \emph{width}. In~\cite{KM19}, the emphasis is put on another version of the explorability game, where tokens can be duplicated, and $|Q|$ is an upper bound for the number of necessary tokens. In this work, we will on the contrary focus on non-duplicable tokens. However, some results of~\cite{KM19} still apply here. In particular the following holds:

\begin{thmC}[{\cite[Rem. 6.9]{KM19}}]
Given an NFA $\A$ and an integer $k$ given in unary, it is \EXPTIME-complete to decide whether $\A$ is $k$-explorable (even if we fix $k=|Q_\A|/2$).
\end{thmC}

The upper bound follows from the fact that the $k$-explorability game of an NFA is a safety game of size $|\A|^k\times|\D|$, where $\D$ is a determinization of $\A$. Overall, this game is exponential in $|\A|+k$. The lower bound is obtained by reducing from an \EXPTIME-complete combinatorial game from \cite{SC79}, which generalizes SAT by allowing two players to alternately choose values for variables. This game is encoded into the $k$-explorability game of an NFA $\A$, with $k$ being half the number of states of $\A$.

We aim here at answering a different question:

\begin{defi}[Explorability problem]
    The explorability problem is the question, given a non-deterministic automaton $\A$, of deciding whether it is explorable (i.e., whether there exists $k\in\N$ such that it is $k$-explorable).
\end{defi}

\noindent
Another difference with the width setting from~\cite{KM19} is that here, some automata are explorable and some are not. Explorable automata can be seen as an intermediary model between deterministic and non-deterministic. Since deterministic and non-deterministic have very different expressive powers for each parity index, this naturally brings the question: what is the expressivity of explorable automata for each parity index?

\textbf{Questions:} Is the explorability problem decidable? If yes, what is its complexity? How expressive are explorable automata for each parity index?


\subsection{Links with HD automata}\label{subsec:HDgen}
An automaton $\A$ is History-Deterministic (HD) if and only if it is $1$-explorable, \ie if there is a strategy $\sigma:\Sigma^*\to Q$ resolving the non-determinism based on the word read so far, with the guarantee that the run piloted by this strategy is accepting whenever the input word is in $L(\A)$. See \eg~\cite{BKKS13} for an introduction to HD automata.

We will give here additional and stronger links between explorable and HD automata. In this part, we will mainly be interested in automata on infinite words.

\subsubsection{Explorability in terms of HDness}\label{subsubsec:Ak_HD}

Similarly to~\cite[Lem 3.5]{KM19}, we can express the $k$-explorability condition as a product automaton being HD.

Let $\A$ be any non-deterministic parity automaton, and $k>0$.

\begin{defi}\label{def:Ak}
We denote by $\A^k$ the union product of $k$ copies of $\A$, i.e. the states are $Q^k$, and $\A^k$ accepts if one of its copies follows an accepting run.
The acceptance condition of $\A^k$ is therefore the disjunction of $k$ parity conditions.
\end{defi}

\begin{lem}\label{lem:Ak_HD}
$\A$ is $k$-explorable if and only if $\A^k$ is HD.
\end{lem}

\begin{proof}
Winning strategies for Determinizer in the $k$-explorability game of $\A$ are in bijection with winning strategies of Determinizer in the $1$-explorability game of $\A^k$.
\end{proof}

%% file: automata/Ak_nonexpl.tex
\begin{tikzpicture}[->,auto,node distance=1cm, every state/.style={minimum size=5pt}, initial text=]

  \node[initial, state] (A) {$p_0$};
  \node[state] (B) [below left=.9cm and 2cm of A] {$p_1$};
  \node[state] (C) [right =of B ] {$p_2$};
  \node (D) [right =of C] {$\dots$};
  \node[state] (E) [right =of D ] {$p_k$};
  \node[state,accepting] (F) [below= 2.1cm of A] {$p_f$};
    \node (labe) [below =.2cm of F] {Explorable $\A_k$};

  \path (A) edge node[above] {$a$} (B);
  \path (A) edge node[right] {$a$} (C);
  \path (A) edge node {$a$} (E);

  \path (B) edge node[left] {$a_1$} (F);
  \path (C) edge node[right] {$a_2$} (F);
  \path (E) edge node {$a_k$} (F);


  \node[initial, state,accepting, right=2.5cm of E] (p) {$q_0$};
  \node[state,accepting] (q) [above right = of p] {$q_1$};
  \node[state,accepting] (r) [below right = of p] {$q_2$};
  \node[state] (s) [below right = of q] {$q_3$};
  \node (labn) [below =.2cm of r] {Non-explorable $\C$};

  \path (p) edge [bend left=15] node[left] {$\Sigma$} (q);
  \path (p) edge [bend right=15] node[left] {$\Sigma$} (r);
  \path (q) edge [bend left=15] node[right] {$a$} (p);
  \path (r) edge [bend right=15] node[right] {$b$} (p);
  \path (q) edge  node[left] {$b$} (s);
  \path (r) edge  node[left] {$a$} (s);
  \path (s)  edge [loop above] node {$\Sigma$} ();

\end{tikzpicture}

%% file: automata/ex_split_exp.tex
\scalebox{.9}{
\begin{tikzpicture}[->,auto,node distance=2cm, every state/.style={minimum size=5pt}, initial text=]

  \node[initial, state] (A0) {$q_0$};
  
  \node[state] (B1) [above right of = A0] {$p_1$};
  \node[state] (C1) [below right of = A0] {$r_1$};
  \node[state] (A1) [below right of = B1] {$q_1$};
  
  \node[state] (B2) [above right of = A1] {$p_2$};
  \node[state] (C2) [below right of = A1] {$r_2$};
  \node[state] (A2) [below right of = B2] {$q_2$};
  
  
  \node (B3) [above right of = A2] {};
  \node (C3) [below right of = A2] {};
  
  \node (dots) [right = 1.5cm of A2] {$\dots$};
  
    \node[state] (Akm) [right = 1.5cm of dots] {$q_{k-1}$};
  
  \node (Bkm) [above left of = Akm] {};
  \node (Ckm) [below left of = Akm] {};

  \node[state] (Bk) [above right of = Akm] {$p_k$};
  \node[state] (Ck) [below right of = Akm] {$r_k$};
  \node[state, accepting] (Ak) [below right of = Bk] {$q_k$};
  

  \path (A0) edge node {$\Sigma$} (B1);
  \path (A0) edge node[below left] {$\Sigma$} (C1);
  
  \path (B1) edge node {$a$} (A1);
  \path (C1) edge node[below right] {$b$} (A1);
  
  \path (A1) edge node {$\Sigma$} (B2);
  \path (A1) edge node[below left] {$\Sigma$} (C2);

  \path (B2) edge node {$a$} (A2);
  \path (C2) edge node[below right] {$b$} (A2);
 
  \path (A2) edge[gray,dashed] (B3);
  \path (A2) edge[gray,dashed] (C3);
  
  \path (Bkm) edge[gray,dashed] (Akm);
  \path (Ckm) edge[gray,dashed] (Akm);
  
  \path (Akm) edge node {$\Sigma$} (Bk);
  \path (Akm) edge node[below left] {$\Sigma$} (Ck);
  
  \path (Bk) edge node {$a$} (Ak);
  \path (Ck) edge node[below right] {$b$} (Ak);
  
\end{tikzpicture}
}

%% file: 02.5_expressivity.tex
\section{Expressivity of explorable automata}

In this section, we ask the following question: what does the parity expressivity hierarchy look like for explorable automata?
Recall that for deterministic automata, this hierarchy is strict, i.e. adding parity ranks allows recognizing more language.
On the contrary, for non-deterministic automata, the hierarchy collapses at the \Buchi level: any $\omega$-regular language can be recognized by a non-deterministic \Buchi automaton.

Finally, let us recall a classical result on expressivity of HD automata. We will also very briefly sketch its proof, as this will be useful in the following.

\begin{lemC}[\cite{BKKS13}]\label{lem:HDdet}
For any parity index $[i,j]$, HD $[i,j]$-automata recognize the same languages as deterministic $[i,j]$-automata.
\end{lemC}
\begin{proof}(Sketch) The HD strategy can always be chosen as using a finite memory $M$. This memory $M$ can be incorporated into the states of the automaton, making it deterministic, without changing its acceptance condition.
\end{proof}

\noindent
We will show in this section that explorable automata have an expressivity that is initially akin to the one of deterministic automata, but surprisingly the parity hierarchy collapses at the level $[1,3]$, i.e. any $\omega$-regular language can be recognized by an explorable $[1,3]$-automaton.

Let us start with the \Buchi case to show some of the behaviours involved.
\begin{lem}\label{lem:Buchidet}
    Languages recognized by explorable \Buchi automata are equal to the languages recognized by deterministic \Buchi automata.
\end{lem}
\begin{proof}
    Notice that the converse inclusion is straightforward: if $\L$ is recognized by a deterministic B\"uchi automata $\D$, then $\D$ is $1$-explorable.

For the direct sense: Let $L$ be a language recognized by a $k$-explorable \Buchi automaton $\A$. We will build a deterministic \Buchi automaton recognizing $L$.
Let $\A^k$ be the union product automaton from \Cref{def:Ak}. By \Cref{lem:Ak_HD}, $\A^k$ is HD.
Moreover, $\A^k$ can easily be turned into a \Buchi automaton: we can exploit the fact the union of finitely many \Buchi conditions is a \Buchi condition, by considering that any transition that is \Buchi on some component is \Buchi globally. This does not change the accepting status of any run, so the resulting \Buchi automaton is still HD, using the same witness strategy as $\A^k$. Thus $L$ is recognized by a HD \Buchi automaton.

As recalled earlier, for any parity index $[i,j]$, HD $[i,j]$-parity automata have same expressivity as deterministic $[i,j]$-automata~\cite{BKKS13}. This is in particular true at the \Buchi level, so there exists a deterministic \Buchi automaton $\D$ recognizing $L$. Notice that in the particular case of \Buchi condition, this deterministic \Buchi automaton can be guaranteed to be polynomial-size with respect to the HD \Buchi automaton~\cite{KS15}, and can also be obtained in \PTIME~\cite{BuchiDet}.
\end{proof}

\noindent
The above proof can easily be adapted to get the following lemma:
\begin{lem}
    Languages recognized by explorable safety (respectively reachability) automata are equal to the languages recognized by deterministic safety (respectively reachability) automata.
\end{lem}

\noindent
Let us first state a general lemma on expressivity of explorable automata with any $\omega$-regular acceptance condition.:
\begin{lem}\label{lem:expl_union}
    Let $\Acc$ be any $\omega$-regular acceptance condition. Then a language $L$ is recognized by an explorable $\Acc$-automaton if and only if $L$ is a finite union of languages recognized by deterministic $\Acc$-automata.
\end{lem}

\begin{proof}
Let us start with the right-to-left direction. Let $L=\bigcup_{1\leq i\leq k} L_i$ , where each $L_i$ is recognized by a deterministic $\Acc$-automaton $\A_i$. We can build an explorable $\Acc$-automaton $\A$ recognizing $L$ by having its initial state branch non-deterministically towards each $\A_i$. This automaton is clearly $k$-explorable, and recognizes $L$.

Assume now that $L$ is recognized by some $k$-explorable $\Acc$-automaton $\A$ with state space $Q$. Let $\sigma$ be a winning strategy in the $k$-token game of $\A$, with finite memory $M$. The existence of such a strategy is guaranteed by the fact that $\omega$-regular languages are closed under union, and any $\omega$-regular game is finite-memory determined. Composing $\A$ with $\sigma$ yields a deterministic automaton $\B$ of state space $Q^k\times M$. Let $\A_i$ be the automaton obtained from $\B$ by considering the $\Acc$-acceptance condition of the $i$-th component. Each $\A_i$ is a deterministic $\Acc$-automaton, and we have $L=\bigcup_{1\leq i\leq k} L(\A_i)$.
\end{proof}

\noindent
The results obtained in the rest of the section are actually corollaries of \Cref{lem:expl_union}, by instantiating $\Acc$ to the different parity conditions: the $[0,2]$ condition is closed under union (yielding \Cref{lem:expr02}), and any parity condition is a union of $[1,3]$ conditions (yielding \Cref{thm:expr13}). However, we will still provide direct proofs for better efficiency of the constructions. Indeed, in the above lemma, each component needs to incorporate a memory structure updating all $k$ tokens, which can be avoided in the direct proofs below.

Let us treat the case of $[0,2]$-automata:
\begin{lem}\label{lem:expr02}
    Languages recognized by explorable $[0,2]$-automata are equal to the languages recognized by deterministic $[0,2]$-automata.
\end{lem}
\begin{proof}
Just as in the proof of \Cref{lem:Buchidet}, the converse inclusion is straightforward, as a deterministic automaton is always explorable.
Let $L$ be a language recognized by some $k$-explorable $[0,2]$-automaton $\A$. Similarly as above, this means that the union product $\A^k$ is HD, and this is witnessed by an HD strategy with finite memory $M$. This allows us to build a deterministic automaton $\B$ with states $Q^k\times M$, where the acceptance condition only depends on the $Q^k$ component, and is a union of $k$ $[0,2]$-conditions.
In order to obtain a deterministic $[0,2]$-automaton, we need to compose $\B$ with a deterministic $[0,2]$-automaton $\C$ on alphabet $\Gamma:=[0,2]^k$, that accepts an infinite word if and only if one of its $k$-components is $[0,2]$-accepting.

Intuitively, the automaton will cycle through the components. Seeing a $1$ in the current component will prompt the automaton to go to the next component while producing a $1$ globally. Seeing a $2$ anywhere causes to produce a $2$ globally (and reset to the first component).
The automaton $\C=(\Gamma,Q_\C,q_0^\C,\delta_\C)$ can be built as follows:
\begin{itemize}
    \item $Q_\C=\{1,\dots,k\}$
    \item $q_0^\C=1$
    \item We directly label the transitions of $\C$ by their parity rank, by giving a transition function $\delta_\C:(Q_\C\times\Gamma)\to (Q_\C\times[0,2])$. It is defined as 
    \[\delta_\C(i,(a_1,\dots,a_k))=\left\{
    \begin{array}{llr}
    (1,2)& \text{ if there is some $j$ with $a_j=2$,}\\
    (i,0)& \text{ otherwise if $a_i=0$}\\
    ((i~\mathrm{mod}~k)+1,1) & \text{ otherwise, i.e. if $a_i=1$.}
    \end{array}\right.\]
\end{itemize}

\noindent
Indeed, it is straightforward to verify that a run of $\C$ is $[0,2]$-accepting if and only if this is the case for one of the $k$ components of its input word. This means that $\C$ deterministically translate the union of $k$ $[0,2]$-conditions into one $[0,2]$-condition.

It now suffices to compose $\B$ with $\C$, i.e. having $\C$ read the ranks output by $\B$, and using the $[0,2]$ acceptance condition of $\C$. This yields a deterministic $[0,2]$-automaton with state space $Q_\B\times\Q_\C$, recognizing $L$.
\end{proof}

\noindent
We finally get to the general case with $[1,3]$-parity:
\begin{thm}\label{thm:expr13}
    Let $L$ be any $\omega$-regular language, there exists an explorable $[1,3]$-automaton that recognizes $L$.
\end{thm}
\begin{proof}
 Without loss of generality, the language $L$ can be recognized by some deterministic $[1,d]$-automaton $\A$ with $d$ even. We will build an explorable $[1,3]$-automaton recognizing $L$.\\
For any given accepting run of $\A$, there is a unique even $l \in [1,d]$ such that $l$ is the biggest priority encountered infinitely often. Conversely, if the run is rejecting, there exists no such even $l$. We will use this property to build an explorable automaton based on $\A$.\\
For $l$ even, we define $\A_l$ as the copy of $\A$ where all priorities $<l$ are replaced with $1$, all priorities $>l$ are replaced with $3$, and all priorities equal to $l$ are replaced with $2$. 
$\A_l$ is thus a deterministic $[1,3]$-automaton. It has the notable property that a word $w$ is accepted by $\A_l$ if and only if it is accepted by $\A$ with highest priority $l$.
Therefore, $\A$ is equivalent to the union of the $\{\A_l~|~l \text{ even }\in [1,d]\}$. We thus build the automaton $\A'$ where the initial state branches non-deterministically via $\varepsilon$-transitions towards all the different $\A_l$ for all even $l\in [1,d]$.
This automaton is non-deterministic, of parity index $[1,3]$, and recognizes exactly the words recognized by $\A$. It is $\frac{d}{2}$-explorable, as this is the maximum number of tokens needed to place one in each $\A_l$ at the start, after which their progression becomes deterministic.
\end{proof}

\noindent
This construction can actually be applied to non-deterministic automata as well, yielding the following result:
\begin{lem}\label{lem:13-expl}
    Let $\A$ be a non-deterministic $[1,d]$-automaton with $d$ even. We can build in \PTIME a $[1,3]$-automaton $\A'$ recognizing $\L(\A)$, such that $\A'$ is explorable if and only if $\A$ is explorable.
\end{lem}
\begin{proof}
       We use the same construction as in the proof of \Cref{thm:expr13} above. This construction is clearly \PTIME in the size of $\A$, and produces an automaton $\A'$ of size $|\A|\cdot\frac{d}2$. Moreover, language equivalence still holds: a word is accepted in $\A$ if and only if it is accepted in at least one of the $\A_l$ if and only if it is accepted in $\A'$. We now need to show the equivalence between explorability of $\A$ and explorability of $\A'$.\\
    If $\A$ is $k$-explorable, then it suffices to initially send $k$ tokens to each copy $\A_l$ and from there use the $k$-explorability strategy in each copy locally. We get that if the input word can be accepted with some token $i$ via the $k$-explorability strategy in $\A$ with highest parity $l$, then the corresponding token is accepting in $\A_l$, and $\A'$ is thus $\frac{kd}{2}$-explorable.\\
    If $\A'$ is $k$-explorable, the behaviour of each of the $k$ tokens can be projected to $\A$, giving a candidate strategy for $k$-explorability of $\A$. This is indeed a valid strategy, as if a token is accepting in $\A_l$, it is also accepting in $\A$. We can conclude that $\A$ is $k$-explorable as well.
\end{proof}

\begin{figure}
    \centering
    \includegraphics[width =.9\textwidth]{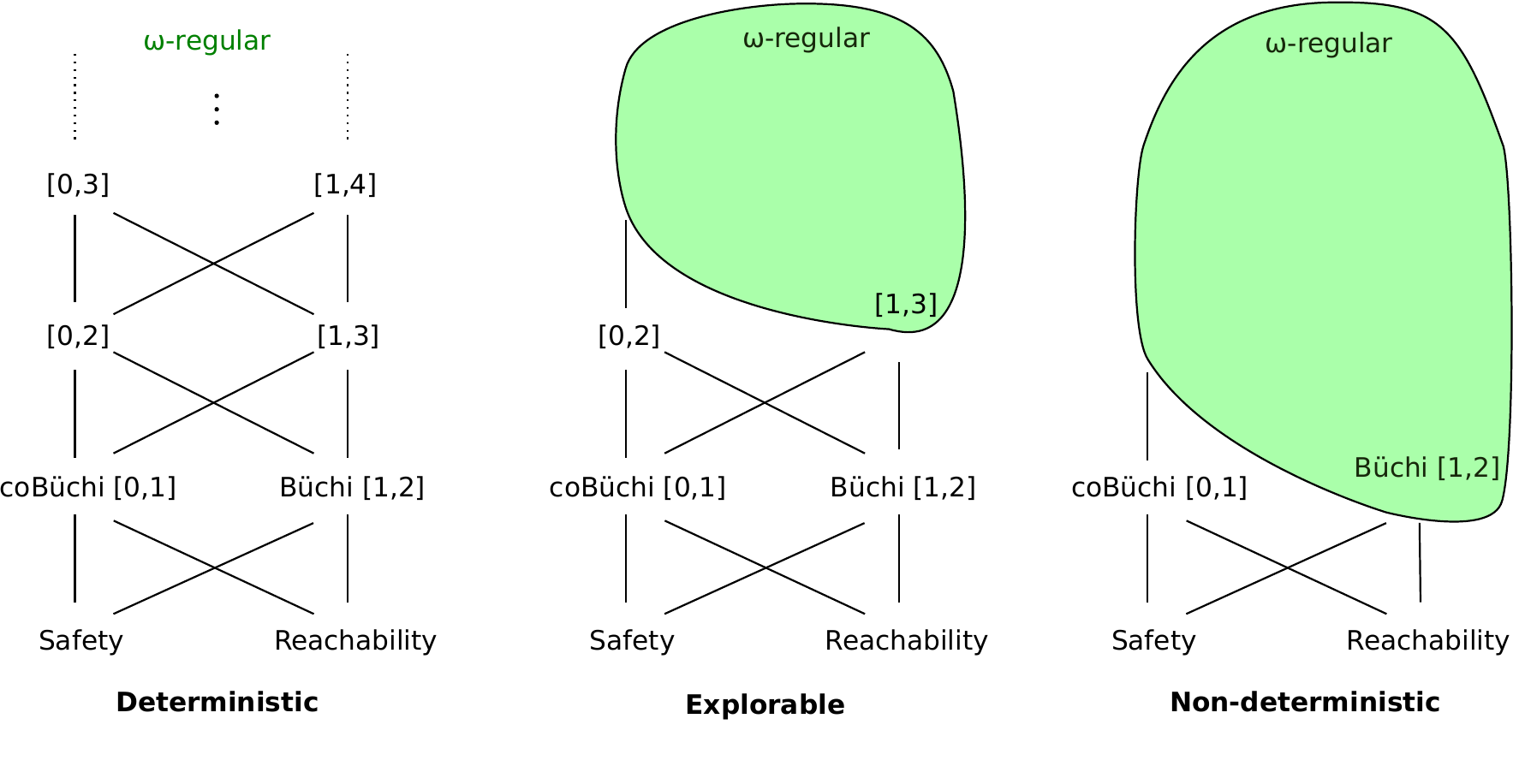}
    \caption{The parity hierarchy of languages recognized by the Deterministic/Explorable/Non-deterministic automata. Recall that the HD hierarchy coincides with the deterministic one. Classes not included in the green collapse region always match their deterministic counterpart.}
    \label{fig:explo-hierarchy}
\end{figure}

\noindent
We thus obtain the hierarchy of languages recognized by explorable automata, represented in Figure \ref{fig:explo-hierarchy}.
This picture will be completed in Section \ref{sec:omBuchi}, where we will show that the hierarchy collapses at the \Buchi level for $\omega$-explorable automata.

%% file: 03_complexity.tex
\section{Decidability and complexity of the explorability problem}

In this section, we prove that the explorability problem is decidable and \EXPTIME-complete for NFAs. We also exhibit an exponential upper-bound for deciding the explorability of $[0,2]$-automata.

We start by showing in Section \ref{sec:blackbox} decidability of the explorability problem for NFAs using the results of~\cite{Bertrand} as a black box. This yields an algorithm in $2$-\EXPTIME. We give in Section \ref{sec:hardness} a polynomial reduction in the other direction, thereby obtaining \EXPTIME-hardness of the NFA explorability problem. To obtain a matching upper bound and show \EXPTIME-completeness, we use again~\cite{Bertrand}, but this time we must ``open the black box'' and dig into the technicalities of their \EXPTIME algorithm while adapting them to our setting.  We do so in  Section \ref{sec:Exptime}, directly treating the more general case of B\"uchi automata.

\subsection{2-\EXPTIME algorithm via a black box reduction}\label{sec:blackbox}

Let us start by recalling the population control problem of~\cite{Bertrand}.
\begin{defi}[$k$-population game]
Given an NFA $\B$ with a distinguished state $\targ\in Q_\B$, and an integer $k\in\N$, the $k$-population game is played similarly to the $k$-explorability game, only the winning condition differs: \controller wins if the game reaches a position where all tokens are in the state $\targ$.
\end{defi}

\noindent The population control problem asks, given $\B$ and $\targ\in Q_\B$,  whether \controller wins the $k$-population game for all $k\in\N$. Notice that this convention is opposite to explorability, where positive instances are defined via a win of \agents.
The population control problem is shown in~\cite{Bertrand} to be \EXPTIME-complete.
We will present here a direct exponential reduction from the explorability problem to the population control problem.

\begin{thm}[Direct reduction to the population control problem]\label{thm:blackbox}
    The NFA explorability problem is decidable in $2$-\EXPTIME.
\end{thm}

\noindent
Let $\A=(\Sigma,Q_\A,q_0^\A,F_\A,\Delta_\A)$ be an NFA. Our goal is to build an exponential NFA $\B$ with a distinguished state $\targ$ such that $(\B,\targ)$ is a negative instance of the population control problem (i.e. the player controlling the tokens can prevent their accumulation in $\targ$) if and only if $\A$ is explorable.

We choose $Q_\B= (Q_\A\times \P(Q_\A))\uplus\{\targ,\bot\}$, where $\targ,\bot$ are fresh sink states.
The alphabet of $\B$ will be $\Sigma_\B=\Sigma\uplus \{\atest\}$, where $\atest$ is a fresh letter.

The initial state of $\B$ is $q_0^\B=(q_0^\A,\{q_0^\A\})$. Notice that we do not need to specify accepting states or priorities in $\B$, as acceptance plays no role in the population control problem.

We finally define the transitions of $\B$ in the following way:
\begin{itemize}
    \item $(p,X)\labelarrow{a}(q,\Delta_\A(X,a))$ if $a\in\Sigma$ and $q\in\Delta_\A(p,a)$,
    \item $(p,X)\labelarrow{\atest}\targ$ if $p\notin F_\A$ and $X\cap F_\A\neq\emptyset$.
    \item $(p,X)\labelarrow{\atest}\bot$ otherwise.
\end{itemize}

\noindent
We aim at proving the following Lemma:
\begin{lem}\label{lem:reduction}
For any $k\in\N$, $\A$ is $k$-explorable if and only if \agents wins the $k$-population game on $(\B,\targ)$.
\end{lem}
\begin{proof}
Notice that as long as letters of $\Sigma$ are played, the second component of states of $\B$ evolves deterministically and keeps track of the set of reachable states in $\A$. Moreover, the letter $\atest$ also acts deterministically on $Q_\B$. Therefore, the only non-determinism to be resolved in $\B$ is how the first component evolves, which amounts to building a run in $\A$. Thus, strategies driving tokens in $\A$ and $\B$ are isomorphic. It now suffices to observe that \controller wins the $k$-population game on $(\B,\targ)$ if and only if he has a strategy allowing to eventually play $\atest$ while all tokens are in a state of the form $(q,X)$ with $q\notin F_\A$ and $X\cap F_\A\neq\emptyset$. This is equivalent to \controller winning the $k$-explorability game of $\A$, since $X\cap F_\A\neq\emptyset$ witnesses that the word played so far is in $L(\A)$.
This concludes the proof that $\A$ is explorable if and only if $(\B,\targ)$ is a negative instance of the population control problem.
\end{proof}

\noindent
Given an NFA $\A$ that we want to test for explorability, it suffices to build $(\B,\targ)$ as above, and use the \EXPTIME algorithm from~\cite{Bertrand} as a black box on $(\B,\targ)$. 
Since $\B$ is of exponential size compared to $\A$, this achieves the proof of \Cref{thm:blackbox}.

\subsection{\EXPTIME-hardness of NFA explorability}\label{sec:hardness}

\begin{thm}\label{thm:hardness}
    The NFA explorability problem is \EXPTIME-hard.
\end{thm}
\begin{proof}
We will perform here an encoding in the converse direction: starting from an instance $(\B,\targ)$ of the population control problem, we build polynomially an NFA $\A$ such that $\A$ is explorable if and only if $(\B,\targ)$ is a negative instance of the population control problem.

It is stated in~\cite{Bertrand} that, without loss of generality, we can assume that $\targ$ is a sink state in $\B$, and we will use this assumption here.

Let $\C$ be the $4$-state automaton of \Cref{ex:explo}, that is non-explorable and accepts all words on alphabet $\Sigma_\C=\{a,b\}$.
Recall that, as an instance of the population control problem, $\B$ does not come with an acceptance condition. We will define its accepting set as $F_\B=Q_\B\setminus\{\targ\}$.

We will take for $\A$ the product automaton $\B\times\C$ on alphabet $\Sigma_A=\Sigma_\B\times\Sigma_\C$, with the union acceptance condition: $\A$ accepts whenever one of its components accepts. The transitions of $\A$ are defined as expected: $(p,p')\labelarrow{a_1,a_2}(q,q')$ in $\A$ whenever $p\labelarrow{a_1}q$ in $\B$ and $p'\labelarrow{a_2}q'$ in $\C$.

Since $L(\C)=(\Sigma_\C)^*$, and $\A$ accepts whenever one of its components accepts, we have $L(\A)=(\Sigma_\A)^*$. The intuition for the role of $\C$ in this construction is the following: it allows us to modify $\B$ in order to accept all words, without interfering with its explorability status.

We claim that for any $k\in\N$, $\A$ is $k$-explorable if and only if \agents wins the $k$-population game on $(\B,\targ)$.

Assume that $\A$ is $k$-explorable, via a strategy $\sigma$. Then \agents can play in the $k$-population game on $(\B,\targ)$ using $\sigma$ as a guide. In order to simulate $\sigma$, one must feed to it letters from $\Sigma_\C$ in addition to letters from $\Sigma_\B$ chosen by \controller. This is done by applying a winning strategy for \controller in the $k$-explorability game of $\C$.
Assume for contradiction that, at some point, this strategy $\sigma$ reaches a position where all tokens are in a state of the form $(\targ,q)$ with $q\in Q_\C$. Since $\targ$ is a sink state, when the play continues it will eventually reach a point where all tokens are in $(\targ,q_3)$, where $q_3$ is the rejecting sink of $\C$. This is because we are playing letters from $\Sigma_\C$ according to a winning strategy for \controller in the $k$-explorability game of $\C$, and this strategy guarantees that all tokens eventually reach $q_3$ in $\C$. But this state $(\targ,q_3)$ is rejecting in $\A$, and $L(\A)=(\Sigma_\A)^*$, so this is a losing position for \agents in the $k$-explorability game of $\A$. Since we assumed $\sigma$ is a winning strategy in this game, we reach a contradiction.
This means that following this strategy $\sigma$ together with an appropriate choice for letters from $\Sigma_\C$, we guarantee that at least one token never reaches the sink state $\targ$ on its $\B$-component. This corresponds to \agents winning in the $k$-population game on $(\B,\targ)$.

Conversely, assume that \agents wins in the $k$-population game on $(\B,\targ)$, via a strategy $\sigma$. The same strategy can be used in the $k$-explorability game of $\A$, by making arbitrary choices on the $\C$ component. As before, this corresponds to a winning strategy in the $k$-explorability game of $\A$, since there is always at least one token with $\B$-component in $F_\B=Q_\B\setminus\{\targ\}$.
This completes the hardness reduction, and thus the proof of \Cref{thm:hardness}.
\end{proof}

\begin{rem}
Using standard arguments, it is straightforward to extend \Cref{thm:hardness} to \EXPTIME-hardness of explorability for automata on infinite words, using any of the acceptance conditions defined in Section \ref{sec:defs}.
\end{rem}

Let us give some intuition on why we can obtain a polynomial reduction in one direction, but did not manage to do so in the other direction. Intuitively, the explorability problem is ``more difficult'' than the population control problem for the following reason. In the population control problem, \controller is allowed to play any letters, and the winning condition just depends on the current position. On the contrary, the winning condition of the $k$-explorability game mentions that the word chosen by \controller must belong to the language of the NFA. In order to verify this, we a priori need to append to the arena an exponential deterministic automaton for this language, and this is what is done in Section \ref{sec:blackbox}. This complicated winning condition is also the source of difficulty in the problem of recognizing HD parity automata.

\subsection{EXPTIME algorithm for [0,2]-explorability}\label{sec:Exptime}
The present work is an extended version of~\cite{HK23}, where the following was proven:
\begin{thm}\label{thm:expBuchi}
The explorability problem can be solved in \EXPTIME for B\"uchi automata (and all simpler conditions: NFA, safety, reachability).
\end{thm}

\noindent
The algorithm was adapted from the \EXPTIME algorithm for the population control problem from~\cite{Bertrand}. We will use a variant of this first algorithm, and thus recall here the main ideas of the latter algorithm, and describe how we adapt it to our setting.

Let $\A$ be an NFA, together with a distinguished state $\targ$.
The idea in~\cite{Bertrand} is to abstract the population game with arbitrary many tokens by a game called the \emph{capacity game}.
This game allows \agents to describe only the support of his set of tokens, \ie the set of states occupied by tokens. Edges that are taken by at least one token are also specified at each step. Thus, it is possible that such a description is not actually realizable with finitely many tokens, e.g. if a state has infinitely many outgoing edges but only finitely many incoming edges throughout the play.
This is captured by a notion of \emph{bounded capacity} for plays in this game: this notion characterizes whether such a play, given as a sequence of supports and egdes, can be instantiated with finitely many tokens.
However, this notion of bounded capacity is not $\omega$-regular. This is why~\cite{Bertrand} introduces the more relaxed notion of \emph{finite capacity}, which is $\omega$-regular, and coincides with bounded capacity in a context of finite-memory strategies.

Thus this property of finite capacity can be verified by a deterministic parity automaton that can be incorporated into the arena. Overall, the capacity game is a parity game that can be won by \controller if and only if $(\A,\targ)$ is a positive instance of the population control problem. Since this parity game has size exponential in $\A$, this yields an \EXPTIME algorithm for the population control problem.

In the present extended version, we will perform some tweaks to this construction, in order to give an \EXPTIME algorithm for deciding explorability of $[0,2]$-automata:

\begin{thm}\label{thm:exp02}
    The explorability problem can be solved in \EXPTIME for $[0,2]$-automata (and all simpler conditions: \Buchi, co-\Buchi, safety, reachability).
    \end{thm}
\noindent
The rest of the section is dedicated to the proof of \Cref{thm:exp02}.

Let us first give a general scheme of the proof, before going to a more formal and detailed description.


We start with a $[0,2]$-automaton $\A$, and want to decide whether it is explorable.

Our aim is to build a parity game $G$ of size exponential in $\A$ such that deciding whether $\A$ is explorable amounts to deciding the winner of $G$. We want $G$ to be an abstraction of the explorability game, where the number of tokens is not explicitly present. First, we need to control that the infinite word played by \controller is in $L(\A)$. This requires to build a deterministic parity automaton $\D$ recognizing $L(\A)$, and incorporate it into the arena. The size of $\D$ is exponential with respect to $\A$, and the number of priorities is polynomial.
We then follow~\cite{Bertrand} and build the \emph{capacity game} augmented with $\D$. In this game, \controller plays a letter at each step, and \agents chooses a subgraph of the run-DAG of $\A$, that intuitively describes all the transitions taken by his tokens. It is then possible to check in an $\omega$-regular way that the run-DAG chosen by \agents can actually be instantiated with finitely many tokens. Thus, this game is a faithful abstraction of the $k$-explorability game for arbitrarily high $k$, if we ignore for now the fact that one token has to follow an accepting run. Then, in order to incorporate this condition as well, we define such a DAG to be $2$-winning if there are infinitely many steps containing a $2$-transition, this is a sufficient condition for one of the token to follow an accepting run. If the DAG is not $2$-winning, we have to additionally check that one token through this run-DAG eventually stops seeing $1$-transitions. Therefore, we add the following feature to the game: \agents has, at all time, a "challenger" that is not supposed to see a $1$-transition. If this constraint is failed infinitely often, the run is said to be $1$-losing. Overall, \agents has to build a run-DAG that is either $2$-winning, or witness via his challenger token that one run through the DAG is not $1$-losing. This allows us to faithfully abstract the $k$-explorability on $[0,2]$-automata for arbitrarily high $k$, via a fixed finite game of size exponential in $\A$ incorporating all these ingredients. This is how we obtain the \EXPTIME algorithm for explorability of $[0,2]$-automata.

We also remark that, as in~\cite{Bertrand}, this construction gives a doubly exponential upper bound on the number of tokens needed to witness explorability. Moreover, the proof from~\cite{Bertrand} that this is tight also stands here.
\medskip

Let us now give a more detailed description of the construction.

We consider a non-deterministic $[0,2]$-automaton $\A = (\Sigma, Q, q_0^\A, \Delta_\A,\rk)$, where $\rk$ is a function $\Delta_A\to[0,2]$ giving the parity rank of each transition. 

Let $\D$ be a deterministic parity automaton for $L(\A)$, that we can obtain via any standard \EXPTIME algorithm. 
We will also make use of the capacity game from~\cite{Bertrand}, in particular let $\T$ be the deterministic parity automaton from\footnote{The structure of this automaton is explicited as the arena of the game $\mathcal{PG}$ in~\cite[Sec 4.3]{Bertrand}. It is described in terms of ``tracking lists'', hence our name $\T$}~\cite[Sec 4.3]{Bertrand}, that accepts a run-DAG if and only if it has infinite capacity. The alphabet of this automaton consists in \emph{transfer graphs} of $\A$, i.e. subsets of $Q\times Q$. 
Both $\D$ and $\T$ have an exponential size and a polynomial number of priorities with respect to the size of $\A$.

\begin{defi}[$\bracket{0,2}$-capacity game]
    The $[0,2]$-capacity game is played in the arena $\P(Q) \times Q \times Q_\D\times Q_\T$, called \emph{$[0,2]$-capacity arena}. Intuitively, The components $\P(Q)$ and $Q_\T$ will be used to simulate the capacity game, while $\P(Q), Q, Q_\D$ will be used as an abstraction of the explorability game. So the component $\P(Q)$, tracking the set of occupied states, is used for both. The $[0,2]$-capacity game is played as follows by \agents and \controller. 
    \begin{itemize}
        \item The starting position is $S_0 = (\{q_0^\A\}, q_0^A, q_0^\D,q_0^\T)$.
        \item At any given step with position $(B, q, q^\D,q^\T)$ with $q\in B$, \controller chooses a letter $a\in\Sigma$, then \agents chooses a transfer graph $G\subseteq \Delta_\A(B,a)$, i.e. a subset of possible $a$-transitions starting from $B$, and a state $q'$ such that $(q,a,q')\in B$.
        \item If $(q,a,q')$ is of priority $1$, then \agents can switch $q'$ to any state in $Im(G)$. This event is recorded as an \textit{elimination}. Else, $q'$ does not change.
        \item The play then moves to the position $(\im(G), q', \delta_\D(q^\D,a),\delta_\T(q^\T,G))$. I.e. the set of tokens is updated to the image of $G$, the state of the challenger is updated to $q'$, and the states of $\D$ and $\T$ are updated deterministically.
    \end{itemize}
    
    \noindent A play can be represented by a sequence $(B_0, q_0, q_0^\D,q_0^\T) \labelarrow{a_1, G_1, e_1} (B_1, q_1, q_1^\D,q_1^\T) \labelarrow{a_2, G_2, e_2}  \ldots$, where $e_i$ is a bit with value $1$ if and only if an elimination took place at step $i$. The state $q_i$ will be called the \textit{challenger}.
    
    We say that \controller wins the play if all the following conditions are simultaneously enforced:
    \begin{itemize}
    \item the run $q_0^\D q_1^\D q_2^\D\dots$ of $\D$ is parity accepting
    \item only finitely many $G_i$ contain 2-transitions (from $\Delta_\A$) 
    \item there is an infinite number of eliminations. 
    \end{itemize}
    Alternatively, \controller also wins if the run $q_0^\T q_1^\T q_2^\T\dots$ of $\T$ is parity accepting, witnessing that the sequence $G_1G_2\dots$ of transfer graphs cannot be instantiated with finitely many tokens.
\end{defi}

The following lemmas will allow us to show that the game functions as intendend, i.e.:
\begin{itemize}
    \item the capacity game still captures the finiteness of number of tokens in this extended construction.
    \item the elimination mechanism is a sound abstraction of the $[0,2]$-parity condition.
\end{itemize}

It thus remains to prove that if Spoiler wins the $[0,2]$-capacity game, if and only if he wins the $k$-explorability game for all $k\in\N$.

Let us begin by an observation that will allow us to prove this result:
\begin{lem}
 The $[0,2]$-capacity game is finite-memory determined.
\end{lem}
\begin{proof}
    The winning condition is a Boolean combination of parity conditions, hence the game is $\omega$-regular, and thus finite-memory determined.
\end{proof}

\noindent
In order to relate the $[0,2]$-capacity game to the $k$-explorability game, we will define the notion of projection of a play.

\begin{defi}[Projection of a play]
    The \emph{support arena} is the $\P(Q)$ component of the $[0,2]$-capacity arena, where \controller plays letters and \agents plays transfer graphs.
    Given a play $S_0 \labelarrow{a_1} S_1 \labelarrow{a_2} S_2 \ldots$ in the $k$-explorability game, the projection of that play in the support arena is the play $B_0 \labelarrow{a_1, G_1} B_1 \labelarrow{a_2, G_2} B_2 \ldots$, where:
    \begin{itemize}
        \item $B_i$ is the support of $S_i$ (states occupied in $S_i$),
        \item $G_{i+1} = \{(S_i(j), S_{i+1}(j)) \mid j \in [ 0, k-1 ]\}$.
    \end{itemize}
    This corresponds to forgetting the multiplicity of tokens and only keeping track of the transitions that are used.
    Any play in the $[0,2]$-capacity game induces a play in the support arena as well, by simply forgetting the challenger and the deterministic components $\D$ and $\T$.
\end{defi}

In the following, we will not recall in detail the notion of capacity or other intricacies of the construction from~\cite{Bertrand}, and will try to use them in a blackbox manner as much as possible.
Since the $[0,2]$-capacity game is an extension of the capacity game from~\cite{Bertrand} with extra components, some results can be readily applied.

In this spirit, combining~\cite[Lem 3.5]{Bertrand} with finite-memory determinacy of the $[0,2]$-capacity game, we obtain the following lemma:

\begin{lem}
    \label{lem:strategy_projection}
    If \agents has a finite-memory winning strategy $\tau$ in the $[0,2]$-capacity game, then he has a strategy $\sigma$ in the $k$-tokens explorability game for some $k$, such that any play consistent with $\sigma$ has its projection consistent with $\tau$.
    Additionnally, it is possible to choose $\sigma$ such that the challenger token in the $[0,2]$-capacity game is always instantiated by a particular token in the $k$-explorability game, that can change at each elimination event in the $[0,2]$-capacity game.
\end{lem}

\begin{proof}
    The strategy $\tau$ guarantees in particular a finite capacity, as \controller wins any play where $\T$ accepts.
    We can therefore directly import the results from~\cite{Bertrand}: the finite memory $m$ of $\tau$ ensures that the capacity of any winning play is actually bounded by a constant depending on $m$. From there,~\cite[Lem 3.5]{Bertrand} ensures that there is a strategy that moves $k$ tokens to realize the transfer graphs of $\tau$, for some $k$ exponential in $m$. This yields overall a doubly exponential upper bound on the number of tokens needed to instantiate the strategy $\tau$.
    Let us now describe how to build $\sigma$ in order to additionally ensure that the challenger token is always instantiated by a particular token in the $k$-explorability game, that can only change at an elimination event. Whenever the challenger is instantiated (at the beginning or at each elimination), $\sigma$ will simply choose the token of minimal index $r$ among those in the new challenger state.
    Then, as all tokens play the same role in the instantiation strategy from~\cite{Bertrand}, it is always possible to have token $r$ follow the challenger path in $\sigma$, until the next elimination event or forever if there is no more elimination. Indeed, we are guaranteed that the wanted transition is always available, as it is part of the current transfer graph $G$.
\end{proof}

\noindent
We can now move to the main results of this section:

\begin{lem}
    \label{lem:capacity-elim_explo_agents}
    If \agents wins the $[0,2]$-capacity game , then he wins the $k$-explorability game for some $k\in\N$.
   
\end{lem}
\begin{proof}
    Since the $[0,2]$-capacity game is finite-memory determined, we can assume that \agents has a finite-memory strategy $\tau$, allowing us to apply \Cref{lem:strategy_projection}. 
    We will describe how to build a strategy for \agents in the $k$-explorability game, for some $k$ given by \Cref{lem:strategy_projection}.
    First of all, we have to lift the actual play in the $k$-explorability game to a play in the $[0,2]$-capacity game. This is done by simply projecting the set of tokens onto their support. The additional choices of challenger token will be given by the strategy $\tau$.
    \agents is therefore able to simulate $\tau$ in this projected $[0,2]$-capacity, against letters played by \controller.
    We can then apply \Cref{lem:strategy_projection} to obtain a strategy $\sigma$ in the $k$-explorability game, instantiating the behaviour of $\tau$ by actual tokens.
    Let us show that the resulting strategy $\sigma$ from \Cref{lem:strategy_projection} is indeed winning in the $k$-explorability game.
  Let $\pi$ be a play of the strategy $\tau$ in the $[0,2]$-capacity game, yielding a corresponding play $\pi'$ of the strategy in the $k$-explorability game.
  If \agents wins $\pi$ by witnessing infinitely many priorities $2$, then one of the $k$ tokens in $\pi'$ will see infinitely many $2$-transitions as well, and therefore $\pi'$ is winning for \agents.
  Otherwise, if \agents wins $\pi$ by preventing the challenger from seeing a $1$-transition, then by the additional property requested of $\sigma$ in \Cref{lem:strategy_projection}, one of the tokens in $\pi'$ will never see a $1$-transition after some point, and therefore $\pi'$ is winning for \agents as well.
  This achieves the proof that \agents has a winning strategy in the $k$-explorability game.
\end{proof}

\noindent
Let us now show the converse direction:

\begin{lem}
    If \agents wins the $k$-explorability game for some $k\in\N$, then he wins the $[0,2]$-capacity game.
\end{lem}
\begin{proof}
Here we do not need any finite-memory determinacy result.
Let $\sigma$ be a winning strategy for \agents in the $k$-explorability game. We will show that this strategy can be lifted to a winning strategy $\tau$ in the $[0,2]$-capacity game.
When playing in the $[0,2]$-capacity game, \agents will keep in memory a corresponding play of the $k$-explorability game, and follow the strategy $\sigma$ in it. Then, to answer to the letters played by \controller, \agents will do the following:
\begin{itemize}
\item project the $k$ tokens to their support, and play the corresponding transfer graph in the $[0,2]$-capacity game
\item for the challenger, simply loop over all tokens: when the current challenger is token $i$ and sees a $1$-transition, an elimination event is witnessed, and the challenger is switched to token $i+1$, looping back to $0$ after $k-1$.
\end{itemize}

\noindent
Since the strategy $\sigma$ is winning in the $k$-explorability game, the corresponding play in the $[0,2]$-capacity game will be winning as well: 
\begin{itemize}
\item the automaton $\T$ will be rejecting as the transfer graphs can be instantiated with a finite number of tokens
\item Either infinitely many $2$ will be seen in the run-DAG, or after some point the challenger will never see a $1$-transition.
\end{itemize}
This achieves the description of a winning strategy $\tau$ for \agents in the $[0,2]$-capacity game.
\end{proof}

\noindent
To conclude and obtain \ref{thm:exp02}, we have to show that the $[0,2]$-capacity game can actually be solved with the wanted complexity.

\begin{thm}
    The $[0,2]$-capacity game can be solved in \EXPTIME.
\end{thm}

\begin{proof}
Let $\G$ be the $[0,2]$-capacity game associated to a $[0,2]$-automaton $\A$. We will show that the winning condition of the game $\G$ for \controller can be seen as a disjunction of parity conditions. Recall that the winning condition for \controller is ($\T$ accepts) or ($\D$ accepts and finitely many $2$ and infinitely many eliminations). Therefore, it is of the form Parity$\vee$(Parity $\wedge$ coB\"uchi $\wedge$ B\"uchi). But we can turn the second disjunct into a parity condition, at the price of some memory. Indeed, coB\"uchi $\wedge$ B\"uchi amounts to a $[1,2,3]$ condition. A condition of the form $[1,2j]$-Parity $\wedge$ $[1,2,3]$-Parity can then be accepted by deterministic $[1,2j+1]$-parity automaton $\D'$ performing the following task: 
\begin{itemize}
\item Whenever a $3$-transition is seen on the second component, produce a rejecting $2j+1$ rank,
\item During sequence of $1$-transitions on the second component, produce a rank of $1$ and remember the highest rank $h$ seen on the first component. This takes $2j$ states.
\item When a $2$-transition is seen on the second component, produce $h$.
\end{itemize}
This automaton $\D'$ has a size polynomial in $\A$, as the number of priorities of $\D$ is polynomial in $\A$.
Thus, by incorporating $\D'$ in the game, we obtain a game $\G'$ equivalent to $\G$, still of size exponential in $\A$, and with winning condition a disjunction of two parity conditions. Such games are studied in~\cite{CHP07}, which gives us an algorithm for solving $\G'$ in time $O(m^{4d}m^2)\frac{(2d)!}{d!^2}$, where $d$ is the number of priorities and $m$ the size of the game.

If we take $n = |\A|$, using the fact that $m = O(2^{\mathrm{poly}(n)})$ and $d=\mathrm{poly}(n)$, we obtain an overall \EXPTIME complexity for solving $\G$.
\end{proof}

\begin{rem} We can also be interested in the number of tokens needed for \agents to witness explorability of an automaton. By inspecting our proof, we can see that we obtain a doubly exponential upper bound. Moreover, we can use the same construction as in~\cite[Prop 6.3]{Bertrand} to show that this is tight, \ie some automata require a doubly exponential number of tokens to witness explorability. It is straightforward to lift this lower bound to the more difficult problem of NFA explorability (or more complex conditions on infinite words), so we do not detail this proof here.
\end{rem}

\subsection{The Parity explorability problem}\label{sec:decidability}

We leave open the decidability of the explorability problem for parity automata beyond index $[0,2]$. 

However, we remark that from Lemma \ref{lem:13-expl}, the only remaining case to be treated is index $[1,3]$. Indeed, for any parity automaton $\A$ with $n$ states and $d$ parity ranks, Lemma \ref{lem:13-expl} allows us to reduce (in polynomial time) explorability of $\A$ to that of an equivalent $[1,3]$-automaton with $\dfrac{nd}{2}$ states.

%% file: 04_infinite_tokens.tex
\section{Explorability with countably many tokens}\label{sec:omega}

In this section, we look at a variant of explorability where we now allow for infinitely many tokens. More precisely, we will redefine the explorability game to allow an arbitrary cardinal for the number of tokens, then consider decidability problems regarding that game. This notion will mainly be interesting for automata on infinite words.

\subsection{Definition and basic results}

The following definition extends the notion of $k$-explorability to non-integer cardinals:

\begin{defi}[$\kappa$-explorability game]
    Consider an automaton $\A$ and a cardinal $\kappa$. The $\kappa$-explorability game on $\A$ is played on the arena $(Q_\A)^\kappa$, between \agents and \controller. They play as follows.
    \begin{itemize}
        \item The initial position is $S_0$ associating $q_0$ to all $\kappa$ tokens.
        \item At step $i$, from position $S_{i-1}$, \controller chooses a letter $a_i\in\Sigma$, and \agents chooses $S_i$ such that for every token $\alpha$, $S_{i-1}(\alpha) \labelarrow{a_i} S_i(\alpha)$ is a transition in $\A$.
    \end{itemize}
    The play is won by \agents if for all $\beta \leq \omega$ such that the word $(a_i)_{1 \leq i < \beta}$ is in $\L(\A)$, there is a token $\alpha \in \kappa$ building an accepting run, meaning that the sequence $(S_i(\alpha))_{i < \beta}$ is an accepting run. Otherwise, the winner is \controller.
\end{defi}

We will say in particular that $\A$ is $\omega$-explorable if \agents wins the game with $\omega$ tokens. We use here the notation $\omega$ for convenience, it should be understood as the countably infinite cardinal $\aleph_0$. We will however explicitly use the fact that such an amount of tokens can be labelled by $\N$, in order to describe strategies for \controller or \agents in the $\omega$-explorability game. 
The following lemma gives a first few results on generalised explorability.

\begin{lem}
    \label{lem:props_omega_explo}
    ~
    \begin{itemize}
        \item \agents wins the explorability game on $\A$ with $|\L(\A)|$ tokens.
        \item $\omega$-explorability is not equivalent to explorability.
        \item There are non $\omega$-explorable safety automata.
    \end{itemize}
\end{lem}

\begin{proof}
    For the first item, a strategy for \agents is to associate a token to each word of $\L(\A)$ and to have it follow an accepting run for that word. 
    Let us add a few details on the cardinality of $L(\A)$. First, a dichotomy result has been shown in~\cite{Niw91} (even in the more general case of infinite trees): if $L(\A)$ is not countable, then it has the cardinality of continuum, and this happens if and only if $L(\A)$ contains a non ultimately periodic word. In this case, we can simply associate a token with every possible run. In the other case where $L(\A)$ is countable, we have to associate an accepting run to each word, and this can be done without needing the Axiom of Countable Choice: a canonical run can be selected (\eg lexicographically minimal).
    
    We now want to prove that there are automata that are $\omega$-explorable but not explorable. One such safety automaton is given in Figure \ref{fig:ex_split_automaton_omega} (left), where the rejecting sink state is omitted. Against any finite number of tokens, \controller has a strategy to eliminate them one by one, by playing $a$ while \agents sends tokens to $q_1$, and $b$ the first time $q_1$ is empty after the play of \agents. On the other hand, with tokens indexed by $\omega$, \agents can keep the token $0$ in $q_0$, and send token $i$ to $q_1$ at step $i$. Those strategies are winning, which proves both non explorability and $\omega$-explorability of the automaton.
    
    The last item is proven by the second example from Figure \ref{fig:ex_split_automaton_omega}. A winning strategy for \controller against $\omega$ tokens consists in labelling the tokens with integers, then targeting each token one by one (first token $0$, then $1$, $2$, \etc). Each token is removed using the correct two-letters sequence ($a$, then $b$ if the token is in $q_1$ or $a$ if it is in $q_2$). With this strategy, every token is removed at some point, even if there might always be tokens in the game.
\end{proof}

\begin{figure}[H]
    \centering
    \input{automata/ex_safety_reach}
    \hspace{1cm}
    \input{automata/ex_split_omega}
    \caption{Two safety automata.\\ Left: $\omega$-explorable but not explorable.\hspace{1cm} Right: not $\omega$-explorable.}
    \label{fig:ex_split_automaton_omega}
\end{figure}

\noindent
The first item of Lemma \ref{lem:props_omega_explo} implies that the $\omega$-explorability game only gets interesting when we look at automata over infinite words: since any language of finite words over a finite alphabet is countable, \agents wins the corresponding $\omega$-explorability game. We will therefore focus on infinite words in the following.

Let us emphasize the following slightly counter-intuitive fact: in the $\omega$-explorability game, it is always possible for \agents to guarantee that infinitely many tokens occupy each currently reachable state. However, even in a safety automaton, this is not enough to win the game, as it does not prevent that each individual token might be eventually ``killed'' at some point.
As the following Lemma shows, this phenomenon does not occur in reachability automata on infinite words.

\begin{lem}
    \label{lem:reachability_always_omega_explo}
    Any reachability automaton is $\omega$-explorable.
\end{lem}

\begin{proof}
    Notice first that although the argument is very similar to the one for finite words, we cannot use exactly the same property: a reachability language can still be uncountable, so using one token per word of the language is not possible.
    
    For every $w \in \Sigma^*$ labelling a finite run $\rho$ leading to an accepting state, \agents can use a single token following $\rho$. This token will accept all words of $w \cdot \Sigma^\omega$. Since $\Sigma^*$ is countable, and all accepted words are accepted after a finite run, we only need countably many such tokens to cover the whole language, hence the result.
    
    Let us give another equally simple view: a winning strategy for \agents in the $\omega$-explorability game is to keep infinitely many tokens in each currently reachable state, as described before the statement of the Lemma. Since acceptance in a reachability automaton is witnessed at a finite time, this strategy is winning.
\end{proof}

\subsection{\EXPTIME algorithm for coB\"uchi automata}
\label{sec:algo_omega_explo}

We already know, from the example of \Cref{fig:ex_split_automaton_omega}, that the result from \Cref{lem:reachability_always_omega_explo} does not hold in the case of safety automata.
For automata which are not automatically $\omega$-explorable, we aim at deciding their $\omega$-explorability status. 
We show the following decidability result on coB\"uchi automata. It holds in particular for safety automata as a subclass of coB\"uchi.

\begin{thm}
    \label{thm:cobuchi_omega_explo_decidable}
    The $\omega$-explorability of coB\"uchi automata is decidable in \EXPTIME.
\end{thm}

\noindent
To prove this result, we will use the following \emph{elimination game}. $\A$ will from here on correspond to a coB\"uchi (complete) automaton. We start by building a deterministic coB\"uchi automaton $\D$ for $L(\A)$ (\eg using the breakpoint construction~\cite{BreakPoint}).
We will assume here that the coB\"uchi condition of $\A$ is state-based to simplify a bit the presentation, but as remarked in the introduction it is straightforward to accomodate transition-based acceptance as well.
\begin{defi}[Elimination game]
    The elimination game is played on the arena $\P(Q_\A) \times Q_\A \times Q_\D$. The two players are named \protector and \eliminator, and the game proceeds as follows, starting in the position $(\{q_0^\A\}, q_0^\A, q_0^\D)$.
    \begin{itemize}
        \item From position $(B, q, p)$ \eliminator chooses a letter $a\in\Sigma$.
        \item If the \emph{challenger} $q$ is not a coB\"uchi state, \protector picks a state $q' \in \Delta_\A(q,a)$.
        \item If the challenger $q$ is a coB\"uchi state, \protector picks any state $q' \in \Delta_\A(B,a)$ as the new challenger. Such an event is called \emph{elimination}.
        \item The play moves to position $(\Delta_\A(B,a), q', \delta_\D(p,a))$.
    \end{itemize}
    Such a play can be written $(B_0, q_0, p_0) \labelarrow{a_1} (B_1, q_1, p_1) \labelarrow{a_2} (B_2, q_2, p_2) \ldots$, and \eliminator wins if infinitely many $q_i$ and finitely many $p_i$ are coB\"uchi states.
\end{defi}

Intuitively, what is happening in this game is that \protector is placing a token that he wants to protect, the challenger, in a reachable state, and \eliminator aims at bringing this challenger to a coB\"uchi state while playing a word of $L(\A)$. If \protector eventually manages to preserve the challenger from elimination on an infinite suffix of the play, he wins.

Notice that this is similar to the technique used for $[0,2]$-explorability, except that in absence of the capacity gadget $\T$, we do not enforce that finitely many tokens are used.

\begin{lem}
    \label{lem:elim_ptime}
    The elimination game is positionally determined and can be solved in polynomial time (in the size of the game).
\end{lem}

\begin{proof}
    To prove this result, we simply need to note that the winning condition is a parity condition of fixed index. If we label the coB\"uchi states $p_i$ of $\D$ with rank $3$, the coB\"uchi states $q_i$ of $\A$ with rank $2$, and the others with $1$, then take the highest rank in $(B_i, q_i, p_i)$ (ignoring $B_i$), \eliminator wins if and only if the highest rank appearing infinitely often is $2$.
    As any parity game with $3$ ranks can be solved in polynomial time~\cite{Zielonka98}, this is enough to get the result. Since parity games are positionnally determined~\cite{EJ91}, the elimination game is as well.
\end{proof}

\noindent
We want to prove the equivalence between this game and the $\omega$-explorability game to obtain Theorem \ref{thm:cobuchi_omega_explo_decidable}.

\begin{lem}
    \label{lem:equiv_elim_omega_explo}
    $\A$ is $\omega$-explorable if and only if \protector wins the elimination game on $\A$.
\end{lem}

\begin{proof}
    First, let us suppose that \eliminator wins the elimination game on $\A$.
    To build a strategy for \controller in the $\omega$-explorability game of $\A$, we first take a function $f: \Nbb \rightarrow \Nbb$ such that for any $n \in \Nbb$, $|f^{-1}(n)|$ is infinite  (for instance we can take $f$ described by the sequence $0,~0,1,~0,1,2,~0,1,2,3, \ldots$).
    The strategy for \controller will focus on sending token $f(0)$, then $f(1)$, then $f(2)$, \etc to a coB\"uchi state.
    Let $\sigma$ be a memoryless winning strategy for \eliminator in the elimination game. \controller will follow this strategy $\sigma$ in the $\omega$-explorability game, by keeping an imaginary play of the elimination game in his memory: $M = \P(Q_\A) \times Q_\A \times Q_\D\times \Nbb$. 
    \begin{itemize}
        \item At first, the memory holds the initial state $(\{q_0^\A\}, q_0^\A, q_0^\D, 0)$, and the current challenger is given by the last component via $f$: it is the token $f(0)$.
        \item From $(B,q,p,n)$ \controller plays in both games the letter $a$ given by $\sigma(B,q,p)$.
        \item Once \agents has played, \controller moves the memory to $(\Delta_\A(B,a), q', \delta_\D(p, a), n)$ where $q'$ is the new position of the token $f(n)$, except if $q$ was a coB\"uchi state, in which case we move to $(\Delta_\A(B,a), q', \delta_\D(p, a), n+1)$ where $q'$ is the new position of the token $f(n+1)$. We then go back to the previous step.
    \end{itemize}
    This strategy builds a play of the elimination game in the memory, that is consistent with $\sigma$.
    We know that $\sigma$ is winning, which implies that the word played is in $\L(\A)$, and that every $n \in \Nbb$ is visited (each elimination increments $n$, and there are infinitely many of those). An elimination happening while the challenger is the token $f(n)$ corresponds, on the exploration game, to that token visiting a coB\"uchi state. Ultimately this means that \agents did not provide any accepting run on any token (by definition of $f$ that visits each index infinitely many times), while \controller did play a word from $\L(\A)$, and therefore \controller wins.

    Let us now consider the situation where \protector wins the elimination game, using some strategy $\tau$. We want to build a winning strategy for \agents in the $\omega$-explorability game. Similarly, this strategy will keep track of a play in the elimination game in its memory. \agents will maintain $\omega$ tokens in any reachable state, while focusing on a particular token which follows the path of the current challenger in the elimination game. When that token visits a coB\"uchi state, we switch to a token in the new challenger state specified by $\tau$.
    
    Since $\tau$ is winning in the elimination game, either the word played by \controller is not in $\L(\A)$, which ensures a win for \agents, or there are no eliminations after some point, meaning that the challenger token at that point never visits another coB\"uchi state, which also implies that \agents wins.
\end{proof}

\noindent
With Lemmas \ref{lem:elim_ptime} and \ref{lem:equiv_elim_omega_explo} we get a proof of Theorem \ref{thm:cobuchi_omega_explo_decidable}, since the elimination game associated to $\A$ is of exponential size and can be built using exponential time.

\subsection{\EXPTIME-hardness of the \texorpdfstring{$\omega$}{omega}-explorability problem}

\begin{thm}\label{thm:om-hard}
    The $\omega$-explorability problem for (any automaton model embedding) safety automata is \EXPTIME-hard.
\end{thm}

\noindent
Before giving the detailed proof of this result, we briefly sketch its main ideas.

\subsubsection{Proof sketch of Theorem \ref{thm:om-hard}}
~
We provide here an intuitive overview of the proof; the full construction is given in Section~\ref{sec:detailed_om_hard}. 
The goal is to reduce the acceptance problem of a \PSPACE alternating Turing machine (ATM)~$\M$ to the $\omega$-explorability problem of a safety automaton~$\A$. 
The computation of an ATM can be viewed as a game between two players, $\exists$ and $\forall$, who respectively aim for acceptance and rejection of the input. 
Each state of the machine belongs to one of the players, who chooses the outgoing transition when in control of that state. 
The input is accepted by the ATM if $\exists$ has a winning strategy in this game. 

The automaton~$\A$ will simulate the configuration of~$\M$ within its set of reachable states, while reading input letters that describe only \emph{local transitions} of~$\M$. 
Our construction is an adaptation of the reduction from~\cite{Bertrand}, which established \EXPTIME-hardness of the NFA population control problem (defined in \Cref{sec:blackbox}). 
We first recall the principles of that reduction and then explain how we modify them to obtain our result.

\paragraph{Alphabet and input words.}
Unlike classical encodings of Turing-machine computations as sequences of configurations, the alphabet here does \emph{not} contain entire configurations, nor even the current control state of $\M$. 
Each letter instead specifies a \emph{local instruction} of the machine:
\begin{itemize}
    \item a transition $t = (q,a,q',a',d)$ of the ATM, and
    \item a tape position~$i$ indicating on which cell this transition is applied.
\end{itemize}
Hence, a word over this alphabet represents a sequence of transitions of~$\M$, describing at each step where the transition occurs, but leaving the global configuration implicit.

\paragraph{How configurations are represented.}
The configuration of~$\M$ is encoded by the \emph{set of states currently reachable} in~$\A$. 
Each state of~$\A$ represents either the content of one tape cell, the current position of the head, or the current control state of the machine. 
Additional states are used for technical purposes, such as recording which player is in control. 
When a letter $(t,i)$ is read, the automaton updates this distributed encoding: it changes the symbol of cell~$i$, moves the head marker, and updates the relevant control component. 
Thus, the global configuration of~$\M$ is not stored in a single state but \emph{distributed} across all reachable states of~$\A$.

\paragraph{Simulation principle.}
During the $\omega$-explorability game, \controller chooses the next input letter~$(t,i)$, specifying which transition of~$\M$ to apply and where, while \agents responds by moving tokens in~$\A$. 
The automaton~$\A$ is constructed so that \controller selects transitions of~$\M$ when in existential control states, through his choice of letters, whereas \agents selects transitions in universal states, by moving to a state $A_t$ of the automaton forcing \controller to play a transition $t$ next.
The component of~$\A$ that updates the configuration of~$\M$ according to these choices is deterministic.

\paragraph{Iterating runs.}
So far, we have described how a single run of~$\M$ is simulated. 
In the reduction, we actually iterate such runs: every time an accepting configuration is reached, \controller may play a special letter that sends some tokens to a designated ``target'' state~$\jail$, and restarts the simulation with the remaining tokens. 
In~\cite{Bertrand}, this mechanism allowed \controller to eliminate tokens one by one, ensuring a win in the population control problem (with finitely many tokens) if and only if he had a winning strategy as~$\exists$ in the ATM game.

\paragraph{Novelty: targeting specific tokens.}
In our setting, merely accumulating tokens in~$\jail$ is insufficient: to win the $\omega$-explorability game, \controller must be able to send \emph{any specific token} to~$\jail$ whenever he reaches an accepting configuration of~$\M$. 
We therefore modify the construction of~\cite{Bertrand} to allow such targeted elimination, taking into account the different components of the automaton $\A$ that may contain the token to eliminate.
Thanks to this refinement, if $\exists$ has a winning strategy in the ATM game, then \controller can use it to repeatedly target and eliminate any token, ensuring victory in the $\omega$-explorability game. 
Conversely, if $\forall$ has a winning strategy, then \agents can always either preserve a ``leader'' token from reaching $\bot$, or reach an accepting sink~$\top$ with some token. Thus \agents win the $\omega$-explorability game.

Overall, $\A$ is $\omega$-explorable if and only if $\M$ rejects its input, yielding the desired reduction and establishing \EXPTIME-hardness.

\subsubsection{Detailed proof of \Cref{thm:om-hard}}\label{sec:detailed_om_hard}
~



We take an alternating Turing machine
\[
\M=(\Sigma_\M, Q_\M, \Delta_\M, q_0^\M, \qaM)
\]
with $Q_\M = Q_\exists \uplus Q_\forall$ and
$\Delta_\M \subseteq Q_\M \times \Sigma_\M \times Q_\M \times \Sigma_\M \times \{L,R\}$.
It can be seen as a game between two players: existential ($\exists$) and universal ($\forall$).
On a given input $w \in (\Sigma_\M)^*$, a play starts in $q_0^\M$, and the owner of the current state resolves non-determinism by picking a transition compatible with the currently read symbol. The input is accepted iff $\exists$ has a winning strategy. We assume $\M$ uses polynomial space $P(n)$ on inputs of length $n$, i.e., winning strategies avoid configurations with tape longer than $P(n)$.

For simplicity, we assume $\Sigma_\M=\{0,1\}$ and $\M$ alternates between $Q_\exists$ and $Q_\forall$, starting in $Q_\exists$ (so transitions alternate $Q_\exists \to Q_\forall \to Q_\exists \to \cdots$). In our reduction, this means the choice of the next transition is given alternately to \controller (for $\exists$) and \agents (for $\forall$).

\paragraph{Safety automaton.}
We build a safety automaton $\A=(Q,\Sigma,q_0,\Delta,\jail)$ where a run is accepting iff it never reaches the rejecting sink $\jail$. Let
\[
Q \;=\; Q_\M \;\uplus\; \Pos \;\uplus\; \Mem \;\uplus\; \Trans \;\uplus\; \{q_0,\store,\jail,\safe\},
\]
with the components:
\[
\begin{aligned}
\Pos &\;=\; [1,P(n)] 
&&\text{(current head position)}\\
\Mem &\;=\; \{\,m_{b,i} \mid b\in\{0,1\},\, i\in[1,P(n)]\,\}
&&\text{(tape content at cell $i$ is $b$)}\\
\Trans &\;=\; \{ \Evestate \} \;\cup\; \{ \Adamstate_t \mid t\in\Delta_\M \}
&&\text{(alternation gadget: players' choices)}
\end{aligned}
\]
and alphabet
\[
\begin{aligned}
\Sigma \;=\; 
&\{\, a_{t,p} \mid t\in\Delta_\M,\; p\in[1,P(n)] \,\}
\;\uplus\;
\{\texttt{init}, \texttt{end}\}\\
&\;\uplus\;
\{\texttt{check}_q \mid q\in Q_\M\}
\;\uplus\;
\{\texttt{check}_{b,i} \mid b\in\{0,1\},\, i\in[1,P(n)]\}.
\end{aligned}
\]

Here $\bot$ is a rejecting sink state: a run is accepting if and only if it never reaches $\bot$.
 


Let us give the intuition for the role of each state of $\A$. First, the states in $Q_\M$, $\Pos$ and $\Mem$ are used to keep track of the configuration of $\M$, as described in Lemma \ref{lem:token_runs}. Those in $\Trans$ are used to simulate the choices of $\exists$ and $\forall$ (played by \controller and \agents respectively). The state \store keeps tokens safe for the remaining of a run when \controller decides to ignore their transition choice. The sinks $\safe$ and $\jail$ are respectively the one \controller must avoid at all cost, and the one in which he wants to send every token eventually.

We now define the transitions in $\Delta$. The states $\safe$ and $\jail$ are both sinks ($\safe$ accepting and $\jail$ rejecting).
We then describe all transitions labelled by the letter $a_{t,p}$ with $p \in \Pos$ and $t = (q, b, q', b', d)\in\Delta_\M$, where $q$ and $q'$ are the starting and destination states of $t$, $b$ and $b'$ are the letters read and written at the current head position, and $d \in \{L,R\}$ is the direction taken by the head. We give below the list of transitions induced by a letter $a_{t,p}$, grouped by purpose:

\noindent\textbf{Update the configuration of $\M$:}
\begin{itemize}
    \item $q \rightarrow q'$.~~\textit{(update control state)}
        \item $p \rightarrow p'$ with $p' = p+1$ if $d=R$, or $p-1$ if $d=L$. Go to $\safe$ if $p' \notin [ 1, P(n)]$.~~\textit{(update head position)}
        \item $m_{b,p} \rightarrow m_{b',p}$, and $m_{b'',p''} \rightarrow m_{b'',p''}$ for any $b''$ and any $p'' \neq p$.~~\textit{(update tape content)}
\end{itemize}
\textbf{Manage alternation between players:}
\begin{itemize}
        \item $\Evestate \rightarrow \Adamstate_{t'}$ for any transition $t'$.~~\textit{(switch to \agents' turn)}
        \item $\Adamstate_t \rightarrow \Evestate$.~~\textit{(switch to \controller's turn)}
\end{itemize}
\textbf{Punish invalid choices:}
\begin{itemize}
        \item $q'' \rightarrow \safe$ for any $q'' \neq q$.~~\textit{(punish invalid control state)}
        \item $m_{\neg b,p} \rightarrow \safe$, where $\neg b$ is not $b$.~~\textit{(punish invalid tape content)}
        \item $p' \rightarrow \safe$ for any $p' \neq p$.~~\textit{(punish invalid head position)}
\end{itemize}
\textbf{Store tokens of non-matching transitions:}
\begin{itemize}
\item $\Adamstate_{t'} \rightarrow \store$ for any transitions $t' \neq t$.
\end{itemize}

\begin{figure}
        \centering
        \begin{tikzpicture}

    \node (Q) {$Q_\M$};
    \node[right=3cm of Q] (Pos) {$\Pos$};
    \node[right=3cm of Pos] (Mem) {$\Mem$};
    \node[right=3cm of Mem] (Trans) {$\Trans$};

    \node[state,below left=.4cm and .1cm of Q] (q) {$q$};
    \node[state, right=of q] (q') {$q'$};
    \node[state,below=2cm of Q] (q'') {$q''$};

    \node[state,below left=.4cm and .1cm of Pos] (p) {$p$};
    \node[state, right=of p] (p') {$p'$};
    \node[state,below=2cm of Pos] (p'') {$p''$};

    \node[state,below left=.4cm and .1cm of Mem] (m) {$m_{b,p}$};
    \node[state, right=of m] (m') {$m_{b',p}$};
    \node[state,below=2cm of Mem] (m'') {$m_{\neg b,p}$};

    \node[state,below left=.4cm and .1cm of Trans] (E) {$\Evestate$};
    \node[state, right=of E] (A) {$\Adamstate_t$};
    \node[state, below=of A] (A') {$\Adamstate_{t'}$};

    \node[state, below=of p''] (safe) {$\safe$};
    \node[state, below=2cm of E] (store) {$\store$};

\coordinate (midPos) at ($(q')!0.5!(p)$);
\coordinate (midMem) at ($(p')!0.5!(m)$);
\coordinate (midTrans) at ($(m')!0.5!(E)$);

\draw[dashed] ($(midPos) + (0,1cm)$) -- ($(midPos) - (0,2.5cm)$);
\draw[dashed] ($(midMem) + (0,1cm)$) -- ($(midMem) - (0,2.5cm)$);
\draw[dashed] ($(midTrans) + (0,1cm)$) -- ($(midTrans) - (0,2.5cm)$);

    \path[->] (q) edge (q');
    \path[->] (p) edge (p');
    \path[->] (m) edge (m');
    \path[->] (E) edge[bend left] (A);
    \path[->] (A) edge[bend left] (E);
    \path[->] (E) edge (A');

    \path[->] (q'') edge (safe);
    \path[->] (p'') edge (safe);
    \path[->] (m'') edge (safe);
    \path[->] (A') edge (store);

\end{tikzpicture}
\caption{Transitions for $a_{t,p}$, where $t=(q,b,q',b',d)$, $p'$ is the position at direction $d$ from $p$, and $q''$, $p''$, and $t'$ are different from $q,p,t$ respectively.}\label{fig:transitions}
\end{figure}
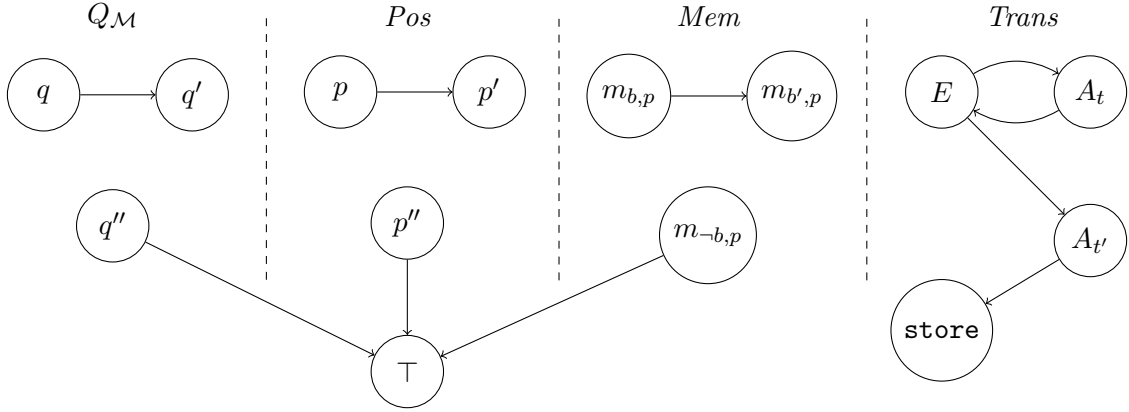

\noindent
These transitions are represented in Figure \ref{fig:transitions}.

Some auxiliary leters will be used to handle other cases, such as the \texttt{check} letters which handle the case where \agents is the one choosing an invalid transition, by sending token to a state $A_t$ when transition $t$ is not compatible with the current configuration of $\M$ stored in components $Q_\M$, $\Pos$ and $\Mem$. 
We list below all transitions induce by other letters of the alphabet (those not of the form $a_{t,p}$):
\begin{itemize}
    \item \textbf{Initialise configuration of $\M$:} \texttt{init} labels transitions from $q_0$ to the states $\Evestate$, $q_0^\M$, and $1\in\Pos$, and also to the states $m_{b,p}$ corresponding to the initial content of the tape, \ie all $m_{b,p}$ such that $b$ is the $p$-th letter of $w$ (or $0$ if $i > |w|$).
    \item \textbf{End current run:} \texttt{end} labels transitions from any non-accepting state of $\M$ to $\safe$, from \store to $q_0$, and from any other state to $\bot$.
    \item \textbf{Punish inconsistent choice of $A_t$ from \agents}:
    \begin{itemize}
    \item $\texttt{check}_q$ labels a transition from $\Adamstate_t$ to $\jail$ for any $t \in \Delta$ starting from $q$. It also labels a transition from $q$ to $\safe$. Any other state is sent back to $q_0$ upon reading $\texttt{check}_q$. Intuitively, playing that letter means that $q$ is not the current state and that any transition starting from $q$ is invalid.
    \item $\texttt{check}_{b,p}$ labels a transition from $\Adamstate_t$ to $\jail$ for any $t \in \Delta$ reading $b$ on the tape. It also labels transitions from any $j \in \Pos \setminus \{p\}$ and from $m_{b,p}$ to $\safe$. Any other state is sent to $q_0$ $\texttt{check}_{b,p}$. Intuitively, playing that letter means that the current head position is $p$, and that its content is not $b$, so any transition reading $b$ is invalid. 
\end{itemize}
\end{itemize}

\noindent
To summarize, the states of $\A$ can be seen as two blocks, apart from $q_0$, $\safe$ and $\jail$: those dealing with the configuration of $\M$ ($Q_\M$, Pos and Mem), and those from the gadget of Figure \ref{fig:gadget_exptime_hard}, dealing with the alternation and non-deterministic choices.

\input{automata/exptime_hard_gadget}
\smallskip

\noindent\textbf{Expected play pattern.}

Let us first describe what a normal play looks like, assuming both players play according to consistent strategies $\sigma_\exists$ and $\sigma_\forall$ in the game of $\M$ on $w$. This gives a simplified view of the expected behaviour of the players in the $\omega$-explorability game on $\A$.
\begin{itemize}
    \item The play starts by \controller playing \texttt{init}, which places tokens in the states representing the initial configuration of $\M$ on $w$, along with a target token (the one with minimal available index) in $\Evestate$.
    \item From there, the players alternate choosing transitions of $\M$:
    \begin{itemize}
    \item when the target token is in $\Evestate$, \controller chooses a transition $t$ according to $\sigma_\exists$ and plays $a_{t,p}$ where $p$ is the current head position;
    \item then \agents moves the token to $\Adamstate_{t'}$  where $t' = \sigma_\forall(\ldots)$, and \controller acknowledges his choice by playing $a_{t',p'}$, where $p'$ is the updated head position, and so on from $\Evestate$ again. Other tokens than the main one may be located in other states $A_{t'',p'}$, corresponding to other valid transitions $t''$ from the target configuration. These ones will be sent to $\store$ when $a_{t',p'}$ is played.
    \end{itemize}
    \item When an accepting configuration of $\M$ is reached, \controller plays \texttt{end}, which sends all non-$\store$ tokens (in particulare the target one) to $\jail$. Tokens in $\store$ are sent back to $q_0$, and the play can restart with another \texttt{init}.
\end{itemize}
The following result provides tools to manipulate the relation between $\A$ and $\M$.

\begin{lem}
    \label{lem:token_runs}
    Let us consider a play of the $\omega$-explorability game on $\A$, that we stop at some point. Suppose that the letters $a_{t,p}$ played since the last \textup{\texttt{init}} are $a_{t_1, p_1}, \ldots, a_{t_k, p_k}$. If $\safe$ is not reachable from $q_0$ with this sequence, then we can define a run $\rho$ of $\M$ on $w$ taking the sequence of transitions $t_1, \ldots, t_k$. The following implications hold:
    
    \centering
    \begin{tabular}{|c|c|}
        \hline
        Token present in & implies that at the end of $\rho$ \\
        \hline \hline
        $q \in Q_\M$ & the current state is $q$\\
        \hline
        $p \in \Pos$ & the head is in position $p$\\
        \hline
        $m_{b,p} \in \Mem$ & the tape contains $b$ at position $p$ \\
        \hline
        $\Evestate$ & it is the turn of $\exists$\\
        \hline
        $\Adamstate_t$ & it is the turn of $\forall$\\
        \hline
    \end{tabular}
\end{lem}

\begin{proof}
    These results are obtained by straightforward induction from the definitions. The unreachability of $\safe$ is used to ensure that only valid transitions are played.
\end{proof}
\noindent
Lemmas \ref{lem:controllerStrat} and \ref{lem:agentStrat} below ensure the correctness of the reduction.
\begin{lem}\label{lem:controllerStrat}
    If $w \in \L(\M)$, then \controller has a winning strategy in the $\omega$-explorability game on $\A$.
\end{lem}
\begin{proof}
There is a winning strategy $\sigma_\exists$ for $\exists$ in the alternating Turing machine game, and \controller will use that strategy in the explorability game to win against $\omega$ tokens. He will consider that the tokens are labelled by integers, and will always target the smallest one that is not already in $\jail$. He proceeds as follows.
\begin{itemize}
    \item \controller plays $\texttt{init}$ from a position where every token is either in $q_0$ or $\jail$. We can assume from here that \agents sends tokens to each possible state, and just add imaginary tokens if he does not. Additionally, if the target token does not go to $\Evestate$, then it means that it is in a deterministic part of the automaton. In this case \controller creates an imaginary target token in $\Evestate$ that will play only valid transitions (we will describe what this means later). Its purpose is to ensure that we actually reach an accepting state of $\M$ to destroy the real target token.
    \item 
    When there are tokens in $\Evestate$, \controller plays letters according to $\sigma_\exists$. More formally, if the letters played since $\texttt{init}$ are $a_{t_1, p_1} \ldots a_{t_i, p_i}$, then \controller plays $a_{t_{i+1},p_{i+1}}$ where $t_{i+1} = \sigma_\exists(t_1, \ldots, t_i)$ and $p_{i+1} = p_i +1$ or $p_i-1$ depending on the head movement in $t_i$.
    \item 
    After such a play, \agents can move tokens to any state $\Adamstate_t$. If he chooses several such states, then \controller will only pay attention to the one containing the current target token.
    \begin{itemize}
        \item If that state $A_t$ corresponds to an invalid transition $t$ (wrong starting state or wrong tape content at the current head position), then \controller plays the corresponding \texttt{check} letter. Formally, if the target token is in $\A_t$, \controller plays $\texttt{check}_q$ if the starting state $q$ of $t$ does not match the current state of the tape (given by Lemma \ref{lem:token_runs}), or $\texttt{check}_{b,p}$ if the current head position is $p$ and does not contain $b$. In both cases, the target token is sent to $\jail$ with no other token reaching $\safe$ (by Lemma \ref{lem:token_runs}). This sends us back to the first step, but with an updated target. 
        \item If $t$ is consistent with the current configuration, then \controller can play the corresponding $a_{t,p}$, where $p$ is the current head position (again, given by Lemma \ref{lem:token_runs}), then go back to the previous step (where there are tokens in $\Evestate$).
    \end{itemize}
    
    \item If the run proceeds normally with consistent transitions, it will eventually reach an accepting state of $\M$ because $\sigma_\exists$ is winning. This corresponds to a stage where \controller can safely play \texttt{end} to send the target token to $\jail$ along with all tokens outside of \store. This sends us back to the first step, but with an updated target. Notice that if there was a virtual target token, we will always reach this event, and send the real target token (located in $Q_\M$ or $\Pos$ or $\Mem$) in $\jail$.
\end{itemize}
This strategy guarantees that after $k$ runs, at least the first $k$ tokens are in state $\jail$, and therefore cannot witness an accepting run. We also know that the final word is accepted by $\A$, because an accepting run can be created by going to the state $\store$ as soon as possible in each factor corresponding to a run of $\M$.
\end{proof}

\begin{lem}\label{lem:agentStrat}
    If $w \notin \L(\M)$, then \agents has a winning strategy in the $\omega$-explorability game on $\A$.
\end{lem}
\begin{proof}
We now assume that there is a winning strategy $\sigma_\forall$ for the universal player in the alternation game on $\M(w)$, and we build a winning strategy for \agents in the $\omega$-explorability game. This strategy is more straightforward than the previous one, as we can focus on the tokens sent to $\Evestate$ (while still populating each state when \texttt{init} is played, but these other tokens follow a deterministic path until the next \texttt{init}). 

\agents will initially choose a specific token, called leader. He then sends $\omega$ tokens to every reachable state when \controller plays \texttt{init}, with the leader going to $\Evestate$. \agents then moves leader according to $\sigma_\forall$, and always keeps every reachable state populated by $\omega$ tokens. \controller cannot send the leader to $\jail$, since the only way to do that would be using the letter \texttt{end}, but this would immediately ensure the win for \agents, as there will always be some token in non-accepting states of $\M$ (because $\sigma_\forall$ is winning), and those tokens would be sent to $\safe$ upon playing \texttt{end}.
This means that \controller has no way to send the leader to $\jail$ without losing the game, and therefore \agents wins.

Note that with that strategy, \controller can still send some tokens to $\jail$ during each run. However, \agents will start the next run with still $\omega$ tokens, including the leader. This is why the choice of a specific leader is important, as this is the witness that an accepting run will be built by $\agents$.

\end{proof}

\noindent
This proves that the automaton $\A$ created from $\M$ and $w$ (using polynomial time) is $\omega$-explorable if and only if $\M$ rejects $w$.
This completes the proof, since the acceptance problem is \EXPTIME-hard for alternating Turing machines using polynomial space.

\subsection{\texorpdfstring{\Buchi}{Buchi} case, or the general case}\label{sec:omBuchi}

Surprisingly, compared to the situation with a finite number of tokens, in the $\omega$-explorability case, the expressivity hierarchy collapses as early as the \Buchi case, as $\omega$-explorable \Buchi automata can recognize all $\omega$-regular languages. We can even build in \PTIME a \Buchi automaton whose $\omega$-explorability is equivalent to the one of an input parity automaton.

\begin{thm}\label{thm:omBuchi}
    Let $\A$ be a parity automaton. We can build in \PTIME a \Buchi automaton $\B$ recognizing $\L(\A)$, such that $\B$ is $\omega$-explorable if and only if $\A$ is $\omega$-explorable.
\end{thm}
\begin{proof}
Intuitively, the idea is to build an automaton that will make a case disjunction over the different even parities $l$ of $\A$, and will ensure that the run never encounters any priority $>l$ after some time. This is a classical way to turn a parity automaton into a \Buchi one.
We just need to ensure that the non-determinism introduced in this construction preserve $\omega$-explorability.

Let us first describe formally the construction.
We define, for $l$ even parity of $\A$, a copy $\A_{l}$ of $\A$ where all transitions of priority $<l$ become non-\Buchi transitions, all transitions of priority $l$ become \Buchi transitions, and all transitions of priority $>l$ are rerouted towards a rejecting sink state $\bot$.
The automaton $\A_{l}$ recognizes the language of words of $\L(\A)$ which can be accepted in $\A$ with infinitely many priorities $l$ and never encounter any priority $>l$.\\
We define $\A'$ as a copy of $\A$ where the rank of all transitions is changed to $1$ (i.e. non-\Buchi).

The automaton $\B$ will simply be the union of $\A'$ and of all the $\A_{l}$: the run starts in $\A'$, then can non-deterministically jump to any $\A_{l}$ at any time, keeping the local state coherent. The transitions of $\B$ are those of $\A',\A_l$, plus transition of the form $p'\trans{a}q_l$ with $p'\in \A'$ and $q_l\in \A_l$ for some $l$, corresponding to a transition $p\trans{a}q$ in the original automaton $\A$, regardless of priorities. 


It is clear that $L(\B)=L(\A)$: an accepting run in $\B$ must jump to some $\A_l$ at some point, and from there witness that the word is accepted in $\A$ with priority $l$. Conversely, a $l$-accepting run in $\A$ can be simulated in $\B$ by jumping to the corresponding $\A_l$ after the last priority $>l$ is encountered.\\

We will now show that $\B$ is $\omega$-explorable if and only if $\A$ is $\omega$-explorable.
\begin{itemize}[align=left]
     \item[$\implies$] If $\B$ is $\omega$-explorable, the strategy for Determinizer can simply be copied to $\A$, by projecting states and transitions of $\B$ to $\A$ in the canonical way. When a token follows an accepting run in $\B$, the corresponding token will follow an accepting run in $\A$, so this strategy witnesses $\omega$-explorability of $\A$.

    \item[$\impliedby$] If $\A$ is \om with $\sigma$ winning strategy for Determinizer, we will build a winning strategy $\sigma'$ for Determinizer in the $\omega$-explorability game $\B$.\\
    To do so, we will associate to each token $i$ of $\sigma$ a countable set of tokens $\{t_{i,j,l}\mid j\in\N, l\text{ even parity of }\A\}$.
    The strategy $\sigma'$ will have token $t_{i,j,l}$ will follow the same path as token $i$ in $\sigma$, starting out in copy $\A'$, and jumping in copy $\A_l$ at time $j$.

    Since some token $i$ accepts in $\sigma$, by seeing infinitely many priority $l$, with no priority $>l$ after some time $j$, the token $t_{i,j,l}$ will accept according to $\sigma'$.

    There are still countably many tokens (which can be re-indexed by $\mathbb N$), so this witnesses $\omega$-explorability of $\B$. \qedhere

\end{itemize}
\end{proof}


\noindent
Theorem \ref{thm:omBuchi} gives us the following corollary:
\begin{cor}
 If $\omega$-explorability is decidable for \Buchi automata, then it is decidable for parity automata.
\end{cor}

\noindent
We leave open the decidability of $\omega$-explorability for \Buchi automata.
\smallskip

The expressivity picture is complete for $\omega$-explorable automata: the hierarchy collapses at the \Buchi level, while the coB\"uchi level recognizes only deterministic coB\"uchi languages, as it is the case for non-deterministic automata in general.

%% file: automata/ex_safety_reach.tex
\begin{tikzpicture}[->,auto,node distance=2cm, initial text=]

  \node[initial, state] (A) {$q_0$};
  \node[state] (B) [right of = A] {$q_1$};
  \node[state] (C) [right of = B] {$q_2$};
  \node[draw = none] (E) at (0,-1) {};
  

  \path (A) edge [loop above] node {$a$} (A);
  \path (A) edge node {$a$} (B);
  \path (B) edge node {$b$} (C);
  \path (C) edge [loop above] node {$a,b$} (C);

\end{tikzpicture}

%% file: automata/ex_split_omega.tex
\begin{tikzpicture}[->,auto,node distance=2cm, initial text=]

  \node[initial, state] (A) {$q_0$};
  \node[state] (B) at (2, 0.8) {$q_1$};
  \node[state] (C) at (2, -0.8) {$q_2$};
  

  \path (A) edge [bend left=0, swap] node {$a$} (B);
  \path (A) edge [bend left=0] node {$a$} (C);
  \path (B) edge [bend right=20, swap] node {$a$} (A);
  \path (C) edge [bend left=20] node {$b$} (A);
  
\end{tikzpicture}

%% file: automata/exptime_hard_gadget.tex
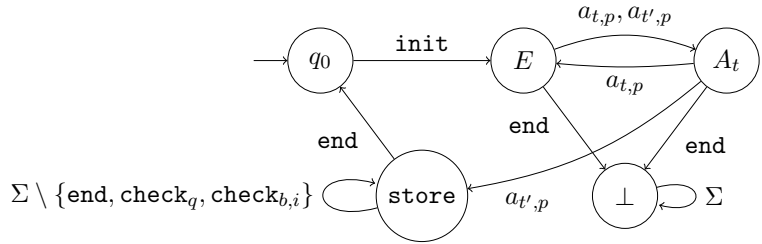
\begin{figure}
    \centering
    \scalebox{.9}{\begin{tikzpicture}[->,auto,node distance=3cm, initial text=]
    
      \node[initial, state] (A) {$q_0$};
      \node[state] (B) [right of = A] {$\Evestate$};
      \node[state] (C) [right of = B] {$\Adamstate_t$};
      \node[state] (E) at (1.5,-2) {$\store$};
      \node[state] (F) at (4.5,-2) {$\jail$};
      
    
      \path (A) edge node {\texttt{init}} (B);
      \path (B) edge [bend left=20] node {$a_{t,p}, a_{t',p}$} (C);
      \path (C) edge [bend left=5] node {$a_{t,p}$} (B);
      \path (C) edge [bend left = 15, pos = 0.9] node {$a_{t',p}$} (E);
      \path (E) edge [loop left] node {$\Sigma \setminus \{\texttt{end}, \texttt{check}_q, \texttt{check}_{b,i}\}$} (E);
      \path (B) edge [pos = .25, swap] node {\texttt{end}} (F);
      \path (C) edge node {\texttt{end}} (F);
      \path (F) edge [loop right] node {$\Sigma$} (F);
      
      \path (E) edge node {\texttt{end}} (A);
      
    \end{tikzpicture}
    }
    \caption{Gadget for simulating the choice of $\forall$ in the alternation (transitions labelled by \texttt{check} are not represented, and $t'$ represents any transition different from $t$).}
    \label{fig:gadget_exptime_hard}
    
\end{figure}

%% file: 05_conclusion.tex
\section*{Conclusion}

We introduced and studied the notions of explorability and $\omega$-explorability, for automata on finite and infinite words.
We showed that these problems are \EXPTIME-complete (and in particular decidable) for $[0,2]$-parity condition in the first case and coB\"uchi condition in the second case.

We leave open the cases of deciding explorability of $[1,3]$-automata and $\omega$-explorability of B\"uchi automata. These correspond to the general case: ($\omega$)-explorability of any parity automaton can be reduced to these cases.

We showed that the original motivation of using explorability to improve the current knowledge on the complexity of the HDness problem for all parity automata cannot be directly achieved, since deciding explorability is at least as hard as HDness. Although this is a negative result, we believe it to be of importance. Moreover, some contexts naturally yield explorable automata, such as~\cite{BL22} where it leads to a \PTIME algorithm deciding the HDness of quantitative LimInf and LimSup automata. More generally, explorability is a natural property in the study of degrees of non-determinism, and this notion could be used in other contexts as a middle ground between deterministic and non-deterministic automata.
We also saw that despite its apparent abstractness, $\omega$-explorability captures a natural property that we believe can be useful in verification: the ability of Spoiler to ``kill'' any run of its choice.
\smallskip

\noindent\textbf{Acknowledgments.} We thank Milla Valnet for her preliminary work on the explorability question, as well as anonymous reviewers for their helpful feedback.